%% file: Dirac-2loops-classification.tex
\documentclass[12pt,aps,prd,a4paper,preprintnumbers,floatfix,nofootinbib,showpacs,superscriptaddress,notitlepage]{revtex4-1}
\pdfoutput=1
\usepackage[english]{babel}
\usepackage{amsmath}
\usepackage{graphicx}
\usepackage{color}
\usepackage{mathrsfs}   %\mathscr{L}
\usepackage{amssymb}
\usepackage[normalem]{ulem} % for striking out text
\usepackage{dcolumn}% Align table columns on decimal point
\usepackage{bm}% bold math
\usepackage{graphicx}
\usepackage[sort&compress]{natbib}
\usepackage{multirow}
\usepackage{rotating}
\usepackage{epstopdf}
\usepackage{diagbox}
\usepackage{booktabs}
\usepackage[utf8]{inputenc}
\usepackage[colorlinks=true,linkcolor=redLinks,citecolor=greenLinks,urlcolor=redLinks, pdfborder={0 0 1}]{hyperref}

\newcommand {\be} {\begin{equation}}
\newcommand {\ee} {\end{equation}}

\definecolor{greenLinks}{rgb}{0, 0.6, 0} 
\definecolor{blueLinks}{rgb}{0, 0, 0.6}
\definecolor{redLinks}{rgb}{0.6, 0, 0}
\definecolor{tempText}{rgb}{0.55, 0.10,0.67}
\definecolor{eprintLinks}{rgb}{0.4, 0.4, 0.4}
\definecolor{journalLinks}{rgb}{0.6, 0, 0}

\def\vev#1{\left\langle #1\right\rangle}

\usepackage{float}

\def\vev#1{\left\langle #1\right\rangle}

\def\21{$\mathrm{SU(2)_L \otimes U(1)_Y}$ }
\def\31{$\mathrm{SU(3)_c \otimes U(1)_Q}$ }
\def\SM{$\mathrm{SU(3)_c \otimes SU(2)_L \otimes U(1)_Y}$ }

\def\3211{$\mathrm{SU(3) \otimes SU(2)_L \otimes U(1)_R \otimes U(1)_{B-L}}$ }
\def\321{$\mathrm{SU(3) \otimes SU(2) \otimes U(1)}$ }
\def\422{$\mathrm{SU(4) \otimes SU(2) \otimes SU(2)_R}$ }
\newcommand {\ignore}[1]{}

\newcommand{\sm}{{Standard Model }}

\def\vev#1{\left\langle #1\right\rangle}

\def\SM{$\mathrm{ SU(3)_C \otimes SU(2)_L \otimes U(1)_Y }$ }
\usepackage{breqn}

\newcommand{\AddrAHEP}{%
  AHEP Group, Institut de F\'isica Corpuscular --
  CSIC-Universitat de València, Parc Cient\'ific de Paterna.\\
 C/ Catedr\'atico Jos\'e Beltr\'an, 2 E-46980 Paterna (Valencia) - SPAIN}
\newcommand{\AddrUNAM}{ {\it Instituto de F\'isica, Universidad Nacional Aut\'onoma de M\'exico, A.P. 20-364, Ciudad de M\'exico 01000, M\'exico.}}

\begin{document}

\title{\large{}Systematic classification of two-loop $d=4$ Dirac neutrino mass models and the Diracness-Dark Matter Stability Connection}
\author{Salvador Centelles Chuli\'a}\email{salcen@ific.uv.es}
\affiliation{\AddrAHEP}
\author{Ricardo Cepedello}\email{ricepe@ific.uv.es}
\affiliation{\AddrAHEP}
\author{Eduardo Peinado}\email{epeinado@fisica.unam.mx}
\affiliation{\AddrUNAM}
\author{Rahul Srivastava}\email{rahulsri@ific.uv.es}
\affiliation{\AddrAHEP}

%%%%%%%%%%%%%%%%%%%%%%%%%%%%%%%%%%%%%%%%%%%%%%%%%%%%%%%%%%%%%%%%%%%%%%%%%%%%%%%%%%%%%%%%%%%%%%%%%%%%%%%%
\begin{abstract}
We provide a complete systematic classification of all two-loop realizations of the dimension four operator for Dirac neutrino masses. Our classification is  multi-layered, starting first  with a classification in terms of  all possible distinct two loop \textit{topologies}.  Then we discuss the possible \textit{diagrams} for each topology. \textit{Model-diagrams} originating from each diagram are then considered. The criterion for \textit{genuineness} is also defined and discussed at length. Finally, as examples, we construct two explicit models which also serve to highlight the intimate connection between the Dirac nature of neutrinos and the stability of dark matter.
\end{abstract}

%%%%%%%%%%%%%%%%%%%%%%%%%%%%%%%%%%%%%%%%%%%%%%%%%%%%%%%%%%%%%%%%%%%%%%%%%%%%%%%%%%%%%%%%%%%%%%%%%%%%%%%%

\maketitle

%%%%%%%%%%%%%%%%%%%%%%%%%%%%%%%%%%%%%%%%%%%%%%%%%%%%%%%%%%%%%%%%%%%%%%%%%%%%%%%%%%%%%%%%%%%%%%%%%%%%%%%%
%%%%%%%%%%%%%%%%%%%%%%%%%%%%%%%%%%%%%%%%%%%%%%%%%%%%%%%%%%%%%%%%%%%%%%%%%%%%%%%%%%%%%%%%%%%%%%%%%%%%%%%%
% Body of the text
\input{Section1_Introduction.tex}
\input{Section2_Classification.tex}

\input{Section3_GeneratingModels.tex}

\input{Section4_Models.tex}

\input{Section5_DM.tex}

\input{Section6_Summary.tex}

%%%%%%%%%%%%%%%%%%%%%%%%%%%%%%%%%%%%%%%%%%%%%%%%%%%%%%%%%%%%%%%%%%%%%%%%%%%%%%%%%%%%%%%%%%%%%%%%%%%%%%%%
%%%%%%%%%%%%%%%%%%%%%%%%%%%%%%%%%%%%%%%%%%%%%%%%%%%%%%%%%%%%%%%%%%%%%%%%%%%%%%%%%%%%%%%%%%%%%%%%%%%%%%%%

\begin{acknowledgments}
Work supported by the Spanish grants SEV-2014-0398, FPA2017-85216-P (AEI/FEDER, UE) and PROMETEO/2018/165 (Generalitat  Valenciana). We thank the support of the Spanish Red Consolider MultiDark FPA2017-90566-REDC. S.C.C. is supported by the grant BES-2016-076643. R.C. is supported by FPU15/03158 (MICINN). E.P. is supported by the grants DGAPA-PAPIIT IN107118 and CONACyT (Mexico) project A1-S-13051 and PIIF-IFUNAM. E.P. would like to thank the group AHEP (IFIC) for their hospitality during his visit there. R.S. would also like to acknowledge the persistent pressure by Martin Hirsch to complete this draft. Without his and Prof. Chuli\'a's constant bullying R.S. might have taken an additional six months to complete it. We thank Renato Fonseca as inspiration for the style of the diagrams.
\end{acknowledgments}

%%%%%%%%%%%%%%%%%%%%%%%%%%%%%%%%%%%%%%%%%%%%%%%%%%%%%%%%%%%%%%%%%%%%%%%%%%%%%%%%%%%%%%%%%%%%%%%%%%%%%%%%
\bigskip
%%%%%%%%%%%%%%%%%%%%%%%%%%%%%%%%%%%%%%%%%%%%%%%%%%%%%%%%%%%%%%%%%%%%%%%%%%%%%%%%%%%%%%%%%%%%%%%%%%%%%%%%
%%%%%%%%%%%%%%%%%%%%%%%%%%%%%%%%%%%%%%%%%%%%%%%%%%%%%%%%%%%%%%%%%%%%%%%%%%%%%%%%%%%%%%%%%%%%%%%%%%%%%%%%
% Appendices
\appendix
\input{Appendix_QuantumNumbers.tex}
\input{Appendix_2loopIntegrals.tex}

%%%%%%%%%%%%%%%%%%%%%%%%%%%%%%%%%%%%%%%%%%%%%%%%%%%%%%%%%%%%%%%%%%%%%%%%%%%%%%%%%%%%%%%%%%%%%%%%%%%%%%%%
%%%%%%%%%%%%%%%%%%%%%%%%%%%%%%%%%%%%%%%%%%%%%%%%%%%%%%%%%%%%%%%%%%%%%%%%%%%%%%%%%%%%%%%%%%%%%%%%%%%%%%%%

\bibliographystyle{BibFiles/utphys.bst}
\bibliography{BibFiles/bibliography}
\end{document}

%% file: Section1_Introduction.tex
%%%%%%%%%%%%%%%%%%%%%%%%%%%%%%%%%%%%%%%%%%%%%%%%%%%%%%%%%%%%%%%%%%%%%%%%%%%%%%%%%%%%%%%%%%%%%%%%%%%%%%%%
%%%%%%%%%%%%%%%%%%%%%%%%%%%%%%%%%%%%%%%%%%%%%%%%%%%%%%%%%%%%%%%%%%%%%%%%%%%%%%%%%%%%%%%%%%%%%%%%%%%%%%%%
\section{Introduction} \label{sec:intro}%%%%%%%%%%%%%%%%%%%%%%%%%%%%%%%%%%%%%%%%%%%%%%%%%%%%%%%%%%%%%%%%
%%%%%%%%%%%%%%%%%%%%%%%%%%%%%%%%%%%%%%%%%%%%%%%%%%%%%%%%%%%%%%%%%%%%%%%%%%%%%%%%%%%%%%%%%%%%%%%%%%%%%%%%
%%%%%%%%%%%%%%%%%%%%%%%%%%%%%%%%%%%%%%%%%%%%%%%%%%%%%%%%%%%%%%%%%%%%%%%%%%%%%%%%%%%%%%%%%%%%%%%%%%%%%%%%

Neutrino masses and the existence of dark matter are the two most important pieces of evidence that the Standard Model is not the final theory of nature. Among the open questions about neutrino physics, probably the most important one is the nature of neutrinos, namely if neutrinos are Dirac or Majorana particles. So far we lack any experimental or observational evidence in favour of one or the other, in spite of the big experimental effort in the last decades, as for instance, in neutrinoless double beta decay \cite{KamLAND-Zen:2016pfg, Agostini:2017iyd, Alduino:2017ehq, Arnold:2016qyg, Albert:2014awa}. 

From a theoretical point of view, Majorana neutrinos have garnered much more attention. Several seesaw \cite{Minkowski:1977sc, Yanagida:1979as, GellMann:1980vs, Mohapatra:1979ia, Schechter:1980gr, Schechter:1981cv, Foot:1988aq} and loop \cite{Zee:1980ai, Zee:1985id, Babu:1988ki, Ma:2006km} mass generation mechanism for Majorana neutrinos have been known for a long time. Furthermore, a systematic classification of all Majorana neutrinos mass mechanisms at a given operator dimensionality and up to certain number of loops also exist in the literature \cite{Ma:1998dn, Babu:2001ex, Bonnet:2009ej, Bonnet:2012kz, Farzan:2012ev, Sierra:2014rxa, Cepedello:2017eqf, Cai:2017jrq, Anamiati:2018cuq, Cepedello:2018rfh, Klein:2019iws}. In contrast, Dirac neutrinos have received relatively little attention. However, in the last few year there has been a renewed interest in looking at mass models for Dirac neutrinos. In this direction, several seesaw mechanisms \cite{Ma:2014qra, Ma:2015mjd, Ma:2015raa, Valle:2016kyz, Chulia:2016ngi, Chulia:2016giq, Reig:2016ewy, CentellesChulia:2017koy, CentellesChulia:2017sgj, Borah:2017leo, Bonilla:2017ekt, Borah:2017dmk, Borah:2018nvu, Ma:2018bow,  Borah:2019bdi} and loop models for Dirac neutrinos \cite{Farzan:2012sa, Okada:2014vla, Bonilla:2016diq, Wang:2016lve, Ma:2017kgb, Wang:2017mcy, Helo:2018bgb, Reig:2018mdk, Han:2018zcn, Kang:2018lyy, Bonilla:2018ynb, Calle:2018ovc, Carvajal:2018ohk, Ma:2019yfo, Bolton:2019bou, Saad:2019bqf, Bonilla:2019hfb, Dasgupta:2019rmf, Jana:2019mez, Enomoto:2019mzl, Ma:2019byo,Restrepo:2019soi} have been recently proposed. Other Dirac neutrino works without a explicit mass mechanism are, for example, \cite{Heeck:2013rpa,Aranda:2013gga,Abbas:2013uqh}. Due to an increasing interest in Dirac neutrino mass models, a classification of such tree-level and one-loop models at dimension 4~\cite{Ma:2016mwh}, dimension 5~\cite{Yao:2018ekp, CentellesChulia:2018gwr} and dimension 6~\cite{Yao:2017vtm, CentellesChulia:2018bkz} have also been considered.

In this paper we will build on these previous works and give a systematic classification of dimension 4 Dirac neutrino mass models at the two-loop level. For the two-loop models to provide the leading order contribution to neutrino masses, one needs to ensure that the dimension 4 tree-level and one-loop contributions are absent. This can happen in a variety of scenarios involving flavour symmetries \cite{Chulia:2016ngi, Chulia:2016giq, CentellesChulia:2017koy} or right-handed neutrinos with chiral lepton number charges \cite{Ma:2014qra, Ma:2015mjd, Ma:2015raa, Bonilla:2018ynb, Bonilla:2019hfb}. Furthermore, one has to ensure that Majorana mass terms are absent at all orders. This can be easily accomplished by, for example, requiring that lepton number (or $B-L$) is conserved exactly \cite{Farzan:2012sa}, or at least an appropriate subgroup of it \cite{Ma:2014qra, Ma:2015mjd, Ma:2015raa, Chulia:2016ngi, Chulia:2016giq, CentellesChulia:2017koy, Hirsch:2017col, Fonseca:2018ehk, CentellesChulia:2018gwr, Bonilla:2018ynb, Calle:2018ovc, Bonilla:2019hfb, Dasgupta:2019rmf}, or also by invoking the presence of additional new symmetries \cite{Ma:2019yfo}.

In this sense, the symmetry protecting the Dirac nature of neutrinos can play a double role and also be used to stabilize a dark matter candidate \cite{Chulia:2016ngi, Chulia:2016giq, CentellesChulia:2017koy, Bonilla:2018ynb, Bonilla:2019hfb}. Moreover, it has recently been shown that all these features, namely Dirac neutrinos, absence of lower order mass terms (tree-level and one-loop) and a stable dark matter can be obtained with chiral, yet anomaly free $B-L$ charges without the need of any other symmetry, either explicit or accidental \cite{Bonilla:2018ynb}.

In this work we consider that neutrino masses are generated from the dimension four operator $\bar{L} \phi^c \nu_R$. Starting from this operator, in general Dirac neutrino mass $m_\nu$ can roughly be written as
\begin{equation} \label{eq:mnu estimate}
    m_\nu \sim C \left( \frac{1}{16 \pi^2} \right)^n \vev{\phi^0},    
\end{equation}
where $n$ is the number of loops needed to generate the Dirac neutrino masses, $\vev{\phi^0} \sim v$ is the SM Higgs vacuum expectation value (vev) and $C$ is a dimensionless constant containing all the information of the couplings involved in the neutrino mass. 
The aim of going to radiative models is that one can explain the smallness of neutrino masses naturally without requiring extremely small couplings. For instance, focusing on the $n = 2$ case and in accordance with the cosmological constraints \cite{Aghanim:2018eyx}, we can take the neutrino masses to be of the order of the atmospheric mass scale $\mathcal{O}(0.05)$ eV \cite{deSalas:2017kay}, which by means of \eqref{eq:mnu estimate} implies couplings of order $0.1$-$0.01$.

The paper is organized as follows: the main body is presented in section~\ref{sec:classification}. We start by classifying all the possible topologies realizing the dimension 4 operator $\bar{L} \phi^c \nu_R$ that can give a dominant Dirac neutrino mass at two-loop level. We then generate all possible diagrams and arrange them in three differentiated classes according to their symmetry and field restrictions. In section~\ref{sec:generating models} we explain how to generate models within each class of diagrams and we then illustrate how the idea can be implemented, discussing in detail two examples in section~\ref{sec:models}. Finally, in section~\ref{sec:dm} we show explicitly the connection between the symmetry that ensures the Dirac nature of neutrinos and the stability of dark matter, closing with a short discussion in the last section. The complete set of tables used to generate models and technical details about the two-loop integrals can be found in the appendices~\ref{sec:appendix qn}~and~\ref{sec:appendix 2loop}, respectively.

%% file: Section2_Classification.tex
%%%%%%%%%%%%%%%%%%%%%%%%%%%%%%%%%%%%%%%%%%%%%%%%%%%%%%%%%%%%%%%%%%%%%%%%%%%%%%%%%%%%%%%%%%%%%%%%%%%%%%%%
%%%%%%%%%%%%%%%%%%%%%%%%%%%%%%%%%%%%%%%%%%%%%%%%%%%%%%%%%%%%%%%%%%%%%%%%%%%%%%%%%%%%%%%%%%%%%%%%%%%%%%%%
\section{Classification: from topologies to models} \label{sec:classification}%%%%%%%%%%%%%%%%%%%%%%%%%%
%%%%%%%%%%%%%%%%%%%%%%%%%%%%%%%%%%%%%%%%%%%%%%%%%%%%%%%%%%%%%%%%%%%%%%%%%%%%%%%%%%%%%%%%%%%%%%%%%%%%%%%%
%%%%%%%%%%%%%%%%%%%%%%%%%%%%%%%%%%%%%%%%%%%%%%%%%%%%%%%%%%%%%%%%%%%%%%%%%%%%%%%%%%%%%%%%%%%%%%%%%%%%%%%%

We start our discussion by first introducing certain key concepts and setting up the notation in generic terms.
Our aim is to consider all possible decompositions of the dimension four operator $\bar{L} \phi^c \nu_R$ at two-loop level for Dirac neutrino masses. 
One of the first key requirements is that the Dirac nature of neutrinos should be protected by some symmetry. This symmetry should remain exact and be such that it forbids the Majorana masses at all loop orders. Such feature can be easily achieved by the global lepton number $U(1)_L$ symmetry already present in the \sm or one of its appropriate unbroken subgroups~\cite{Hirsch:2017col}. However, in general grounds this symmetry protection can in principle have different origins and need not be related with lepton number.

Given an appropriate symmetry protecting the Dirac nature of neutrinos, the next issue is regarding the leading contribution to neutrino masses. Since in this work we are interested in two-loop UV completions of the operator $\bar{L} \phi^c \nu_R$, the tree-level and one-loop contributions should be absent.
This can also happen naturally in many models involving an additional $Z_2$ symmetry \cite{Chulia:2016ngi}, a flavour symmetry \cite{Chulia:2016giq,CentellesChulia:2017koy} or chiral $U(1)_L$
charges for $\nu_R$ \cite{Ma:2014qra, Ma:2015mjd, Ma:2015raa, Bonilla:2018ynb, Bonilla:2019hfb}. 
In fact, it has been recently shown that an appropriate residual subgroup of the lepton number (or equivalently $B-L$ symmetry) alone is enough to guarantee the  Dirac nature of neutrinos to all loops and ensure that the leading contribution to neutrino mass only arises at higher loops \cite{Bonilla:2018ynb}.
Since all the requirements to have Dirac neutrino masses with a leading contribution at two-loops can always be meet, henceforth we will take a different approach and not bother about the details of the symmetries required for an specific model. Instead, we will rather focus on the classification of all such possible models in general.

In this section we begin by looking at how one can systematically organize and analyse all the two-loop realizations of $\bar{L} \phi^c \nu_R$. Before starting with the classification, let us first define a few important concepts which we will use henceforth:
\begin{itemize}
 \item \textbf{Topology}: We define \textit{topologies} as the Feynman diagrams where no Lorentz nature of the involved fields is considered.
 \item  \textbf{Diagram}: We call \textit{diagrams} to those topologies where the fermion and scalar lines are specified.
 \item  \textbf{Model-Diagram}: When the quantum numbers of the internal fields of a diagram are explicitly given, we name them \textit{model-diagrams}.
\end{itemize}

Another key concept is the \textit{genuinity} of a topology or diagram. We identify as \textit{genuine} those model-diagrams (and consequently the topologies or diagrams which generate them) for which the main contribution to the neutrino masses arises at two-loops. On the contrary to the Majorana case \cite{Bonnet:2012kz, Sierra:2014rxa, Cepedello:2018rfh}, for Dirac neutrinos one always needs a symmetry argument to forbid the Yukawa coupling $\bar{L} \phi^c \nu_R$ at tree-level, as discussed before. For this reason, every finite 1 Particle Irreducible (1PI) topology\footnote{Note that, in general, loop integrals have finite and divergent parts. In any consistent renormalization scheme infinite integrals always require a lower order counter term to absorb the infinities, so they are never genuine.} is genuine in our sense providing the correct symmetry and transformations of the fields in order to avoid lower order contributions. 

Although in principle the choice of the symmetries used to forbid tree-level masses and ensuring the Dirac nature of neutrinos is model-dependent, some general conclusions can be given in order to establish a useful classification for model builders.

It is important to clarify that if one imposes a symmetry that forbids the tree-level Dirac mass term, all the realizations of the operator $\bar{L} \phi^c \nu_R (\phi^\dagger \phi)^n$ with $n\in\mathbb{N}$ will automatically vanish. For this reason, one must break such a symmetry, either softly or spontaneously, in order to allow radiative or higher-dimensional Dirac neutrino mass models such that the tree-level term is still absent \cite{Ma:2014qra, Chulia:2016ngi, Bonilla:2018ynb}. Keep in mind that any model with a softly-broken symmetry can be replaced by one with spontaneous breaking by just adding a vev-carrying scalar with the adequate charge in all the soft-breaking terms. Since this would increase the dimensionality of the UV complete Dirac mass operator, we will not study it further.

%%%%%%%%%%%%%%%%%%%%%%%%%%%%%%%%%%%%%%%%%%%%%%%%%%%%%%%%%%%%%%%%%%%%%%%%%%%%%%%%%%%%%%%%%%%%%%%%%%%%%%%%
\subsection{Topologies} \label{subsec:topos}
%%%%%%%%%%%%%%%%%%%%%%%%%%%%%%%%%%%%%%%%%%%%%%%%%%%%%%%%%%%%%%%%%%%%%%%%%%%%%%%%%%%%%%%%%%%%%%%%%%%%%%%%

\begin{figure}
\centering
    \includegraphics[scale=0.7]{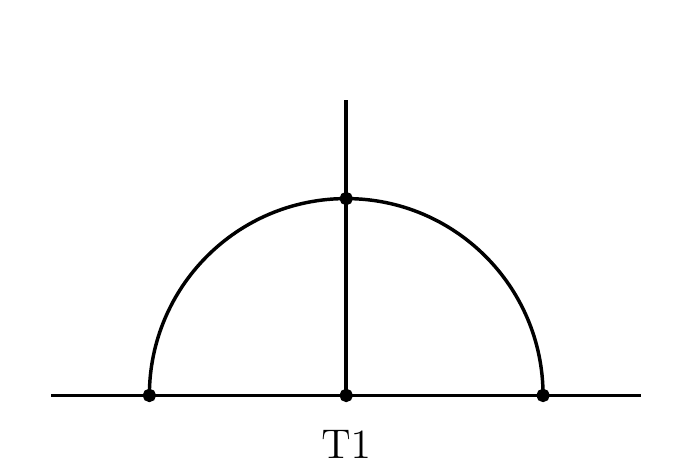}
    \includegraphics[scale=0.7]{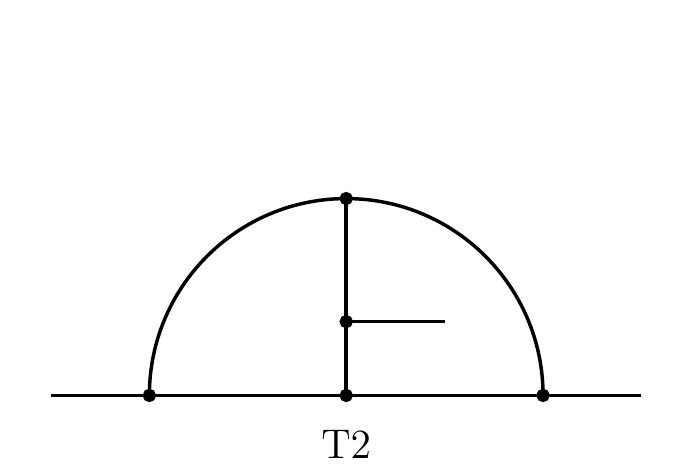}
    \\
    \includegraphics[scale=0.7]{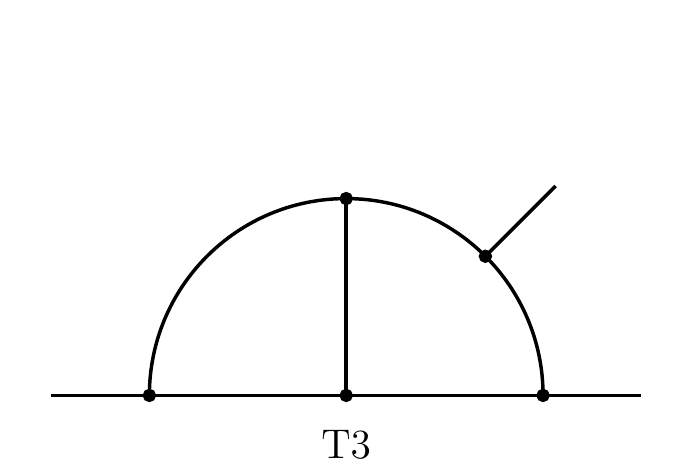}
    \includegraphics[scale=0.7]{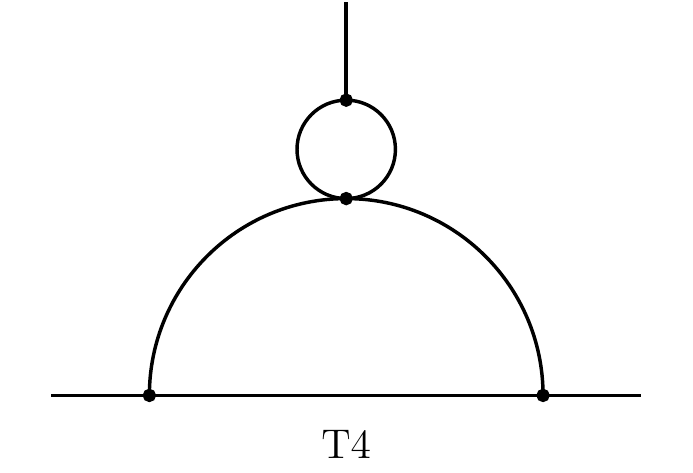}
    \includegraphics[scale=0.7]{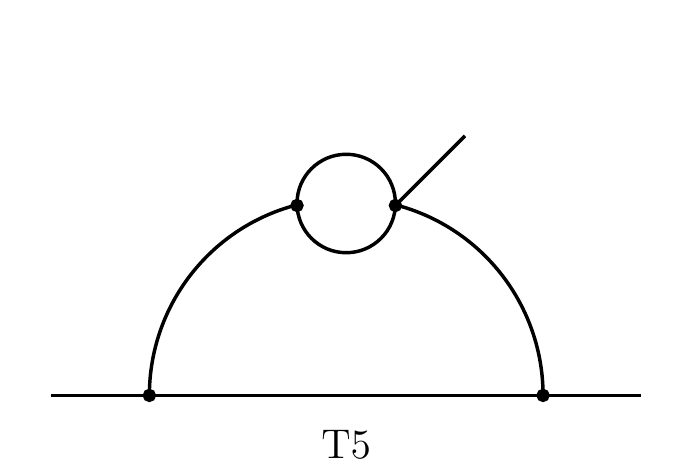}
    
    \caption{\footnotesize Finite two-loops 1PI topologies with 3- and 4-point vertices and three external lines. $T1$ and $T2$ generate in general genuine models. Topologies $T3$, $T4$ and $T5$ are a special kind of genuine topologies, diagrams generated from them are finite but contain a 3-point or 4-point renormalizable vertex which can be in principle reducible, generating a lower order contribution. See the text for details.}
    \label{fig:topologies} 
\end{figure}

We generate all possible connected topologies with two-loops, 3- and 4-point vertices and three external lines. This gives a total number of 70 topologies. 
From these 70 topologies we remove all the topologies corresponding to tadpoles and self-energy diagrams as they always imply infinite parts in the loop integral.
Furthermore, since at the topology level we have not specified the Lorentz nature of the lines, there are also non-renormalizable diagrams, for instance, 3-point vertices with only fermions or 4-point vertices with a fermion insertion.
After removing all these topologies a small set of 5 1PI topologies remains, shown in figure \ref{fig:topologies}.

These 5 1PI topologies can be divided into two differentiated sets of topologies. Topologies $T3$, $T4$ and $T5$ contain an internal loop that can be compressed to a 3-point vertex, whereas the two remaining topologies $T1$ and $T2$ do not. One can argue that the latter are genuine, while the former are corrections to their corresponding one-loop topologies, as they contain a loop realization of a renormalizable vertex. Nevertheless, there are various ways to address this reducibility, leading to differentiated classes at the diagram level.

\begin{figure}
\centering
    \includegraphics[width=\textwidth]{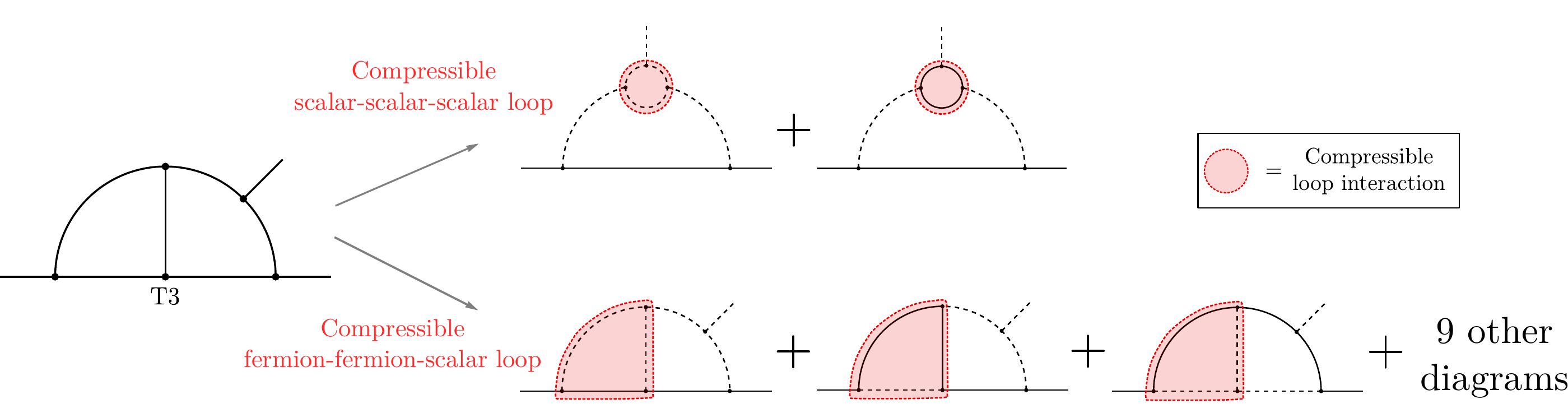}
    \caption{\footnotesize Set of renormalizable diagrams generated from topology $T3$. We found that the topology is not genuine at the diagram level as all the diagrams contain a reducible 3-point loop vertex, coloured in red for each case. The same happens with $T4$ and $T5$. Among the diagrams two differentiated sets are given regarding the type of reducible loop vertex. This will be important at the model-diagram level, in order to promote this a priori non-genuine diagrams to genuine in sections \ref{subsec:class 2} and \ref{subsec:class 3}.}
    \label{fig:diagrams T3}
\end{figure}

As an example, one can construct all possible diagrams of topology $T3$ and see whether there is a reducible (renormalizable) loop interaction. In figure \ref{fig:diagrams T3}, we can see the procedure to follow. Given a topology one should introduce scalar and fermion lines such that there are two external fermions and one external scalar, as required for the diagram to generate effectively $\bar{L} \phi^c \nu_R$. Then, one should check if any diagram contains a loop realization of a renormalizable vertex (marked in red in figure \ref{fig:diagrams T3}). If so, a priori, the diagram is not genuine. If all of the diagrams generated from a certain topology are not genuine, then that topology too is not genuine \cite{Bonnet:2012kz, Sierra:2014rxa, Cepedello:2018rfh}. This is the standard way to address genuinity. In general, one can classify topologies as genuine checking reducibility on any loop at the diagram level. Nevertheless, loopholes to this procedure can be found and we will exploit them in the subsequent sections.

We now move to the construction and classification of genuine diagrams. From the set of 5 topologies, a total of 18 diagrams can be built with two external fermionic lines, one external scalar and containing only renormalizable vertices, depicted in Figs.~\ref{fig:diagrams 1},~\ref{fig:diagrams 2}~and~\ref{fig:diagrams 3}.  One immediately finds three classes among the 18 diagrams by looking at the compressibility of one-loop vertices in the diagram, as well as the Lorentz nature of these vertices.

Note that all these conclusions so far are derived taking into account only fermions and scalars, but they can be directly generalized to include vectors. No new topology or diagram appear if vectors are considered. To extend our classification to vectors one just has to replace one or more scalars with vectors, provided that the resulting diagram is still renormalizable\footnote{Note that some vertices with vectors cannot be built such as vector-vector-vector-scalar.}.

%%%%%%%%%%%%%%%%%%%%%%%%%%%%%%%%%%%%%%%%%%%%%%%%%%%%%%%%%%%%%%%%%%%%%%%%%
\subsection{Completely genuine diagrams} \label{subsec:class 1}
%%%%%%%%%%%%%%%%%%%%%%%%%%%%%%%%%%%%%%%%%%%%%%%%%%%%%%%%%%%%%%%%%%%%%%%%%

\begin{figure}[h]
\centering
    \includegraphics[scale=0.7]{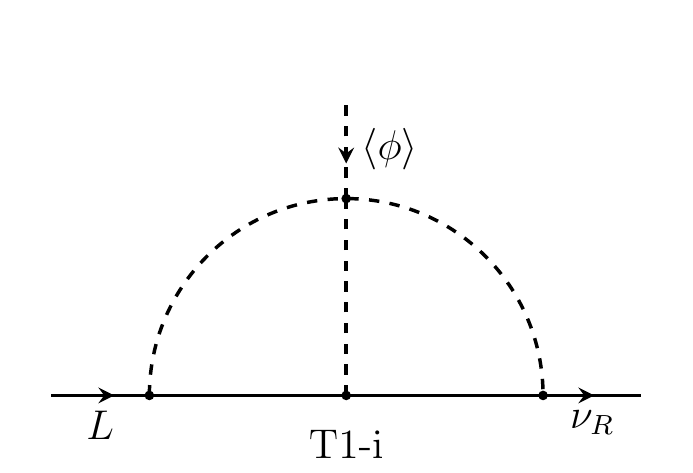}
    \includegraphics[scale=0.7]{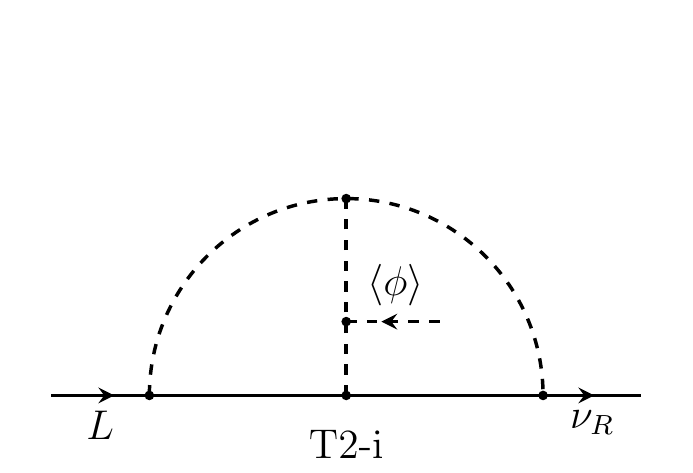}
    \includegraphics[scale=0.7]{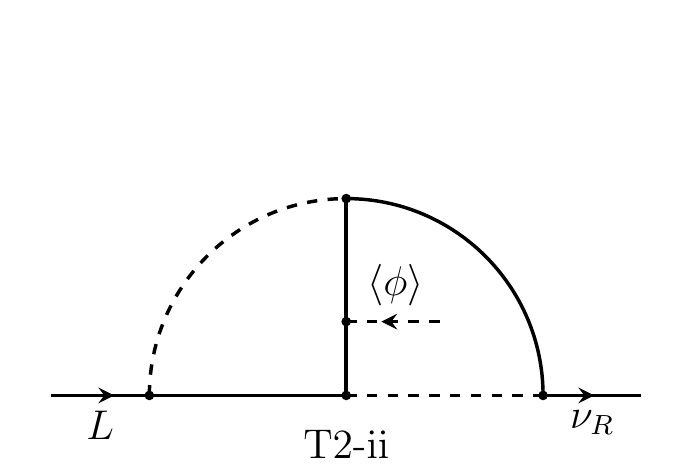}
    
    \caption{\footnotesize Set of diagrams which, in general, do not generate a lower order contribution, i.e. they contain no reducible 3- or 4-point renormalizable vertex. However, the tree-level Dirac mass term should be forbidden by some symmetry.}
    \label{fig:diagrams 1} 
\end{figure}

The topologies $T1$ and $T2$ do not contain compressible renormalizable sub-parts. All the diagrams generated from these topologies will be genuine in our sense. The complete list of diagrams for these two topologies is given in figure \ref{fig:diagrams 1}.

Note that the concept of genuinity stated above implies that these diagrams will generate at least one model at leading two-loop order. Of course, if the symmetries and particle content are not well chosen, one can find specific sets of fields for these diagrams that generate a lower order contribution.

As said before, diagrams generated from the rest of topologies ($T3$, $T4$, $T5$), will not be straightforwardly genuine, due to the presence of a compressible 3-point vertex. Nevertheless, genuinity can be addressed in various ways and general conclusions can be drawn, as we will discuss in the next sections.

%%%%%%%%%%%%%%%%%%%%%%%%%%%%%%%%%%%%%%%%%%%%%%%%%%%%%%%%%%%%%%%%%%%%%%%%%%%%%%%%%%%%%%%%%%%%%%%%%
\subsection{Diagrams with a compressible fermion-fermion-scalar vertex} \label{subsec:class 2}%%%
%%%%%%%%%%%%%%%%%%%%%%%%%%%%%%%%%%%%%%%%%%%%%%%%%%%%%%%%%%%%%%%%%%%%%%%%%%%%%%%%%%%%%%%%%%%%%%%%%

Naively, we could be tempted to think that, in general, the diagrams of this class are a set of corrections to the corresponding one-loop diagrams obtained by shrinking the fermion-fermion-scalar loop vertex. It is trivial to see that if a renormalizable loop vertex is allowed by the symmetries, so should be the vertex without the loop. This means that a priori this diagrams will not be genuine in general, as they always generate a one-loop diagram. However, there are ways to avoid this issue. One can try to forbid the tree-level vertex which generates the lower order one-loop diagram by adding an extra symmetry to the Standard Model. Then this symmetry can be softly broken to allow the loop vertex leading to a two-loop genuine diagram.

\begin{figure}
\centering
    \includegraphics[scale=0.7]{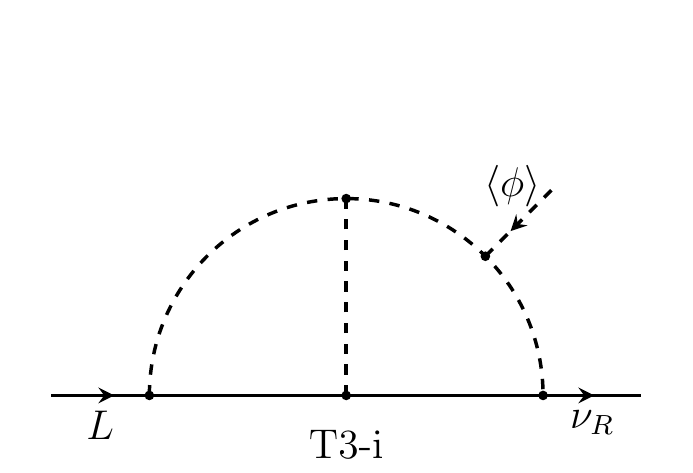}
    \includegraphics[scale=0.7]{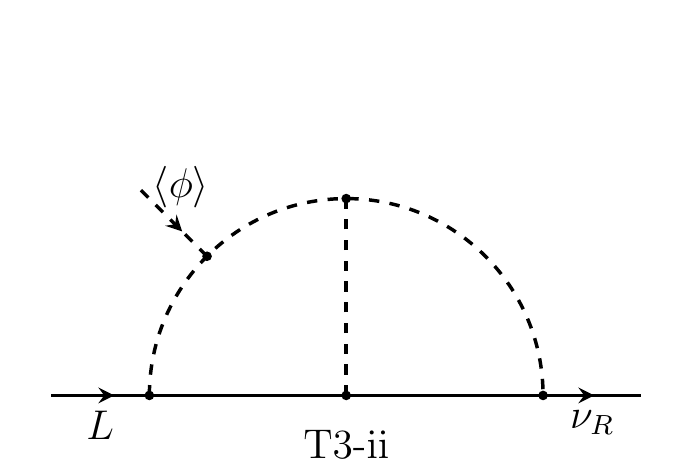}
    \includegraphics[scale=0.7]{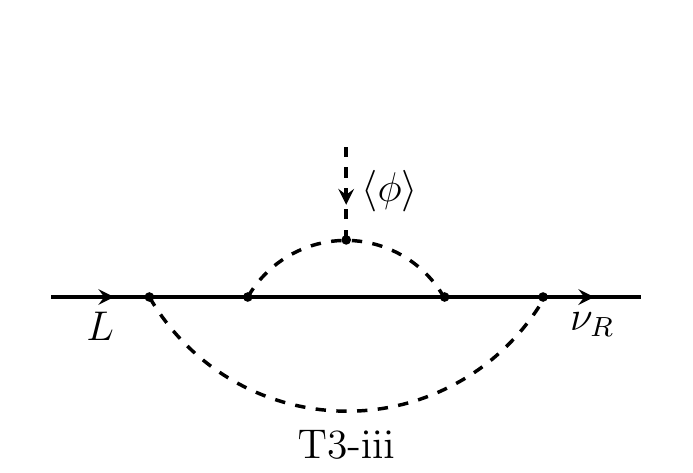}
    \\
    \includegraphics[scale=0.7]{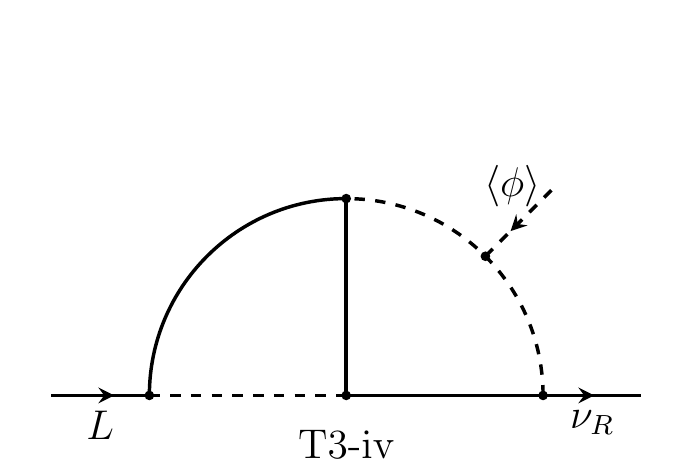}
    \includegraphics[scale=0.7]{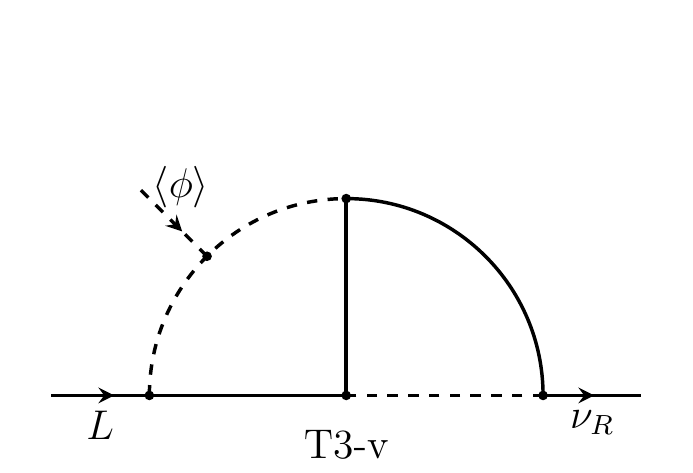}
    \includegraphics[scale=0.7]{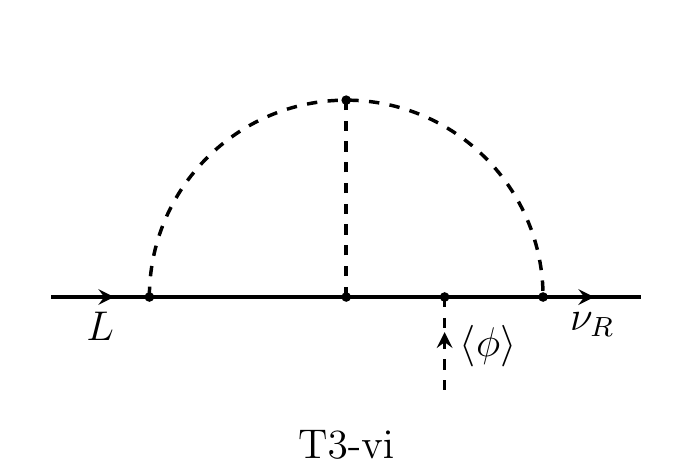}
    \\
    \includegraphics[scale=0.7]{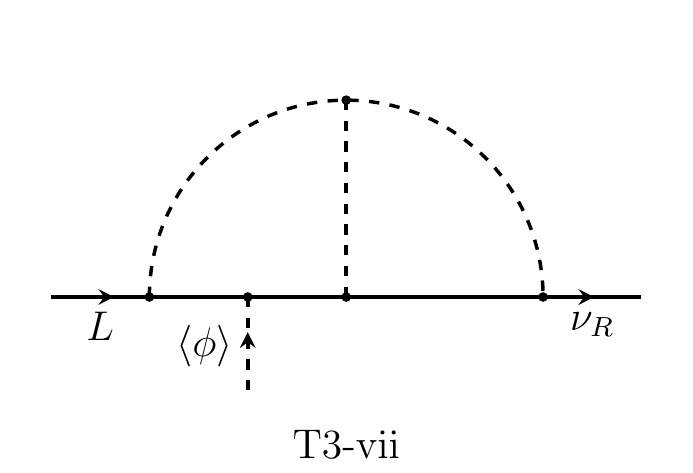}
    \includegraphics[scale=0.7]{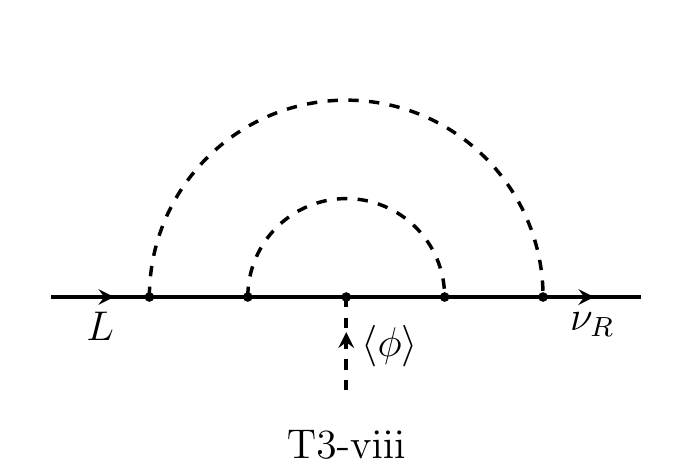}
    \\
    \includegraphics[scale=0.7]{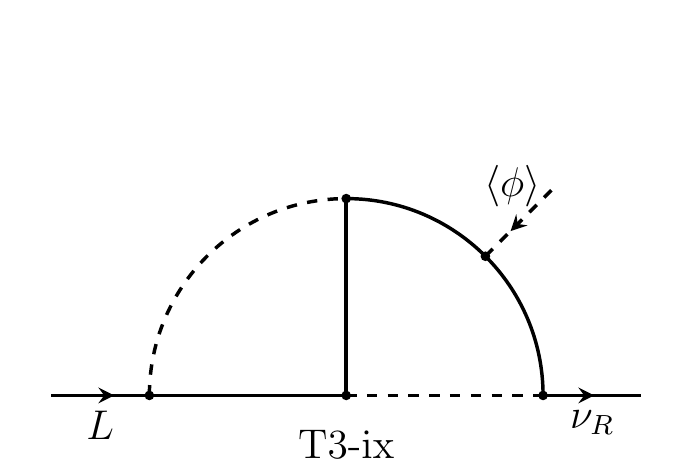}
    \includegraphics[scale=0.7]{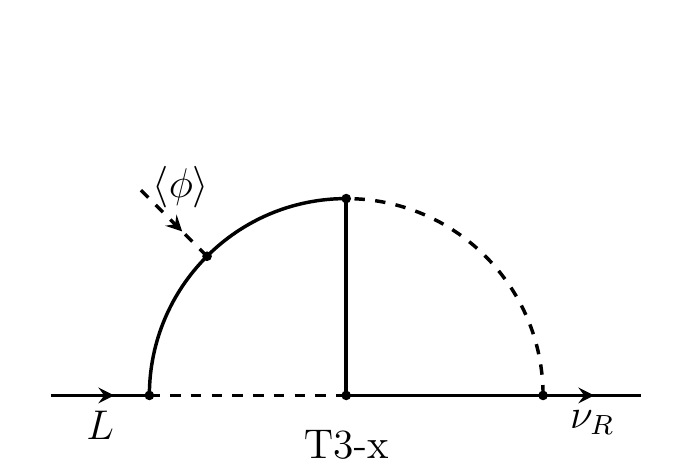}

    \caption{\footnotesize Set of finite diagrams with a compressible fermion-fermion-scalar vertex. These two-loop diagrams can be the dominant contribution to neutrino masses, provided the lower order contributions are forbidden by a softly broken symmetry.  Note that each diagram contains a \textit{soft vertex} which breaks the symmetry forbidding the lower order diagrams.}
    \label{fig:diagrams 2}
\end{figure}

Given the correct extra symmetry and transformation of the fields, one can use the same symmetry which forbids the Dirac tree-level mass term to forbid the tree-level vertex and then break it softly to allow the loop vertex.
Since in this case the forbidden fermion-fermion-scalar coupling is a hard vertex, the tree-level realization is still absent after the soft breaking, while the model remains completely renormalizable. In this sense, the soft breaking term should be contained in a soft coupling or mass participating in the fermion-fermion-scalar loop vertex.
This procedure thus forbids the lower order loop diagram but allows the two-loop diagram making use of the same symmetry that protects Diracness. The diagrams in this class are given in the figure~\ref{fig:diagrams 2}.

It is worth mentioning that there is at least another way to avoid the compressibility of a fermion-fermion-scalar loop vertices. Instead of using a softly broken symmetry, one can force the fermion-fermion-scalar loop vertex to contain a derivative choosing the correct chirality of the fermions. This would make the effective tree-level coupling non-renormalizable (dimension 5 or beyond) \cite{Cepedello:2018rfh}.

%%%%%%%%%%%%%%%%%%%%%%%%%%%%%%%%%%%%%%%%%%%%%%%%%%%%%%%%%%%%%%%%%%%%%%%%%%%%%%%%%%%%%%%%%%%%%%%%
\subsection{Diagrams with a compressible scalar-scalar-scalar vertex} \label{subsec:class 3}%%%%
%%%%%%%%%%%%%%%%%%%%%%%%%%%%%%%%%%%%%%%%%%%%%%%%%%%%%%%%%%%%%%%%%%%%%%%%%%%%%%%%%%%%%%%%%%%%%%%%
 
Analogous to the class of diagrams discussed in Sec.~\ref{subsec:class 2}, the diagrams with compressible three scalar vertex are also, in general, corrections to a one-loop neutrino mass diagrams. Nevertheless, in this case the procedure with softly broken symmetries (or derivatives) does not work. The problem arises due to the fact that a three scalar vertex is a soft term, so the tree-level vertex needs to be included in order to have a consistent renormalizable model. This makes the procedure useless as the one-loop diagram cannot be absent.

\begin{figure}
\centering
    \includegraphics[scale=0.8]{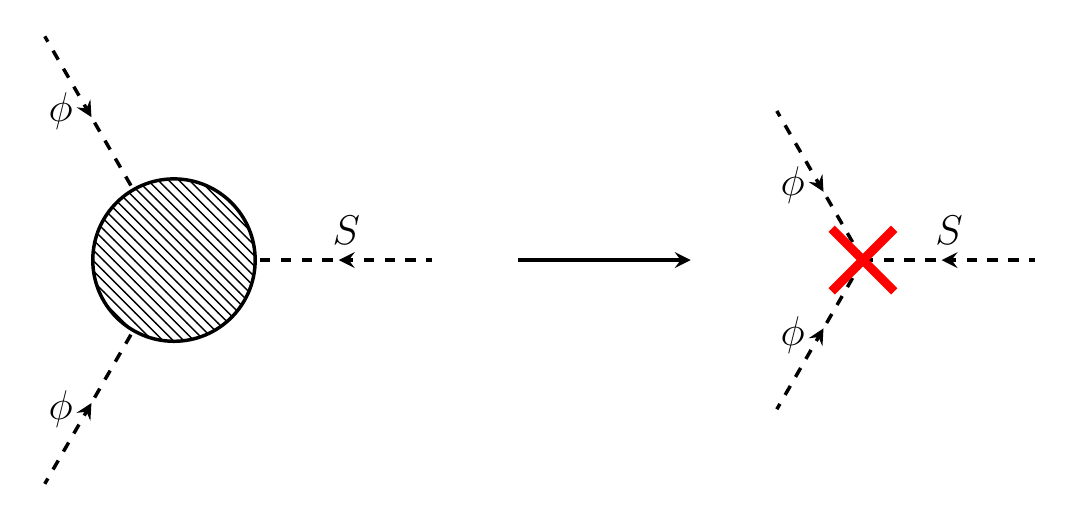}
        
    \caption{\footnotesize For the SM Higgs $\phi$ and the scalar singlet $S$ transforming as $(\mathbf{1},\mathbf{1},-1)$ under the \sm gauge group, the local operator $\phi(x)\phi(x)S(x)$ (right) is automatically zero, as the contraction under $SU(2)_L$ of two doublets to a singlet is antisymmetric. On the contrary, the non-local operator $\phi(x_1)\phi(x_2)S(x_3)$ (left) does not vanish in general, but implies the difference of two diagrams \cite{Cepedello:2018rfh}.}
    \label{fig: example HHS}
\end{figure}

The solution was first pointed out in \cite{Cepedello:2018rfh} where the authors introduce a scalar $S$ transforming as $(\mathbf{1},\mathbf{1},-1)$ under the \SM gauge group. The idea is that the antisymmetric $SU(2)_L$ contractions makes the vertex $\phi \phi S$ exactly zero at tree-level, while the one-loop (non-local) realization of the same operator is in general non zero, see figure~\ref{fig: example HHS}. This can only be applied to scalar-scalar-scalar vertices with just one copy of the Higgs and the new singlet $S$. 
 
\begin{figure}
\centering
    \includegraphics[scale=0.7]{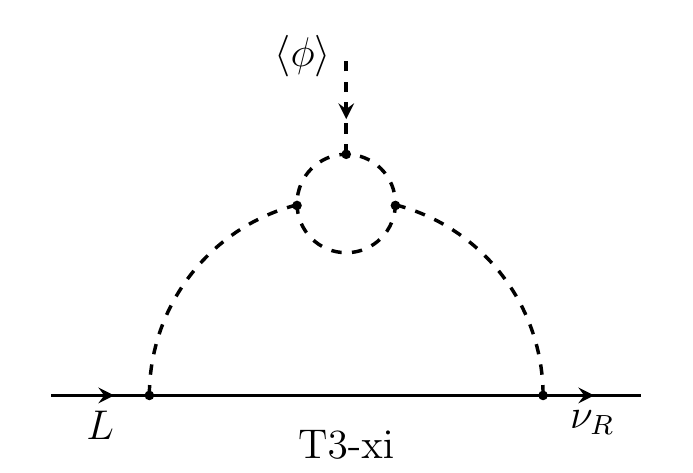}
    \includegraphics[scale=0.7]{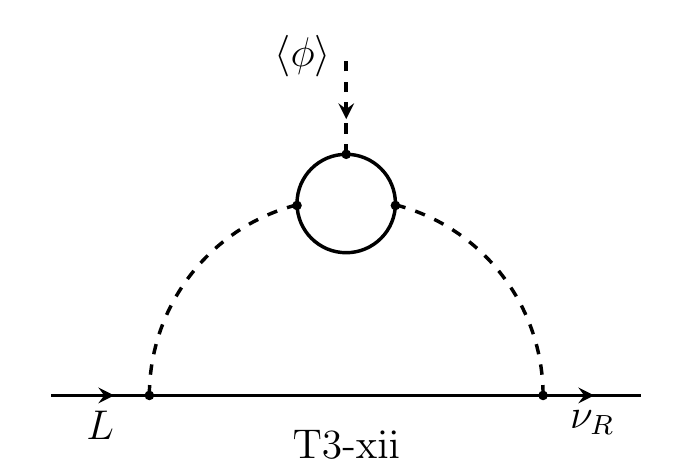}
    \includegraphics[scale=0.7]{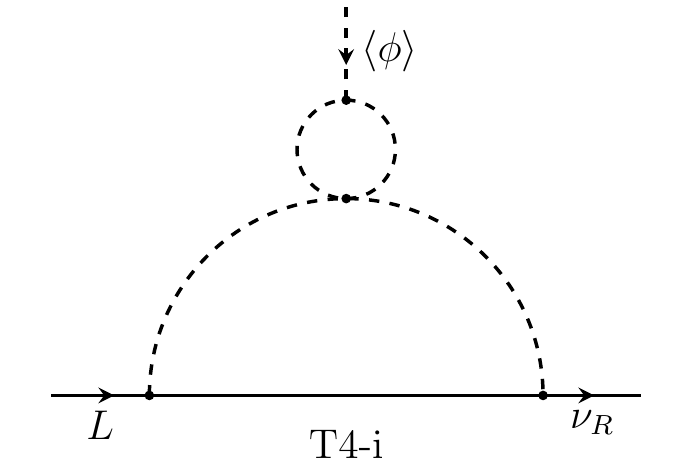}
    \\
    \includegraphics[scale=0.7]{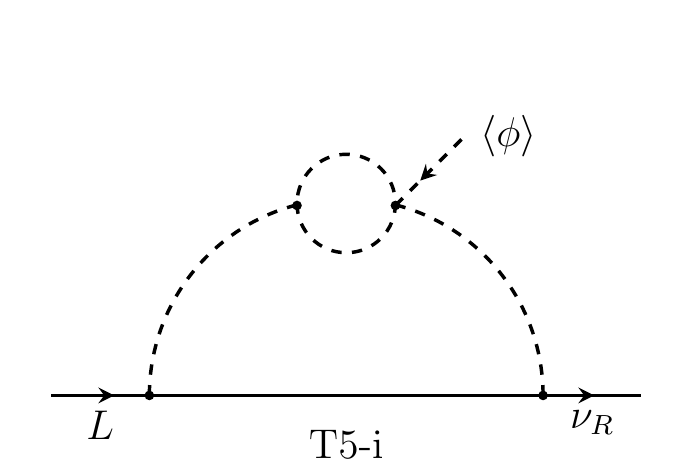}
    \includegraphics[scale=0.7]{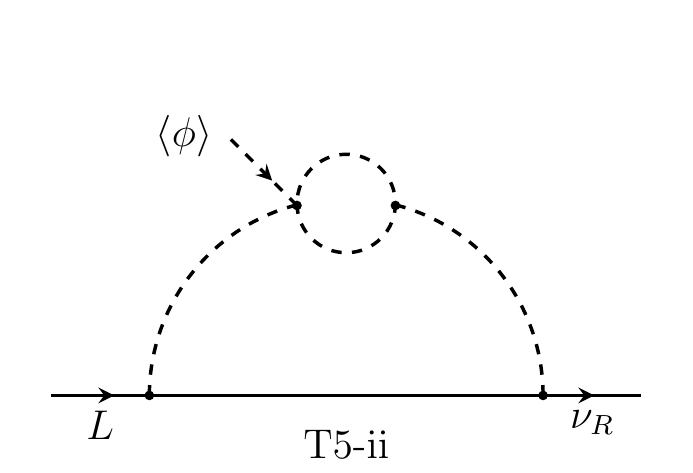}
    
    \caption{\footnotesize Diagrams with a compressible scalar-scalar-scalar vertex. All these diagrams contain a one-loop (i.e. non-local) realisation of a 3-point scalar vertex. In each case, the tree-level vertex $\phi(x)\phi(x)S(x)$ is exactly zero, thanks to the fact that the antisymmetric nature of the $SU(2)_L$ contraction of the two doublets to a singlet. See figure~\ref{fig: example HHS} for details.}
    \label{fig:diagrams 3}
\end{figure}

In figure~\ref{fig:diagrams 3} we give all the diagrams which fall into this class. All the genuine models generated from these diagrams should contain just one Higgs and one scalar $S\equiv(\mathbf{1},\mathbf{1},-1)$. In order to fit the neutrino mass spectra, all such diagrams should also contain at least two copies of a new vector-like pair of fermions with exactly the same \SM charges as those of either $e_R$ or $L$.

In general, the field content of the models generated from this class of diagrams is extremely constrained. Contrary to other two-loop diagrams, here there is only one free choice for the colour, $SU(2)_L$ representation or hypercharge of the particles running in the loops.

%%%%%%%%%%%%%%%%%%%%%%%%%%%%%%%%%%%%%%%%%%%%%%%%%%%%%%%%%%%%%%%%%%%%%%%%%%%%
\subsection{Diagrams in the mass basis} \label{subsec:massdiagrams}%%%%%%%%%
%%%%%%%%%%%%%%%%%%%%%%%%%%%%%%%%%%%%%%%%%%%%%%%%%%%%%%%%%%%%%%%%%%%%%%%%%%%%

\begin{figure}[h]
\centering
    \includegraphics[scale=0.7]{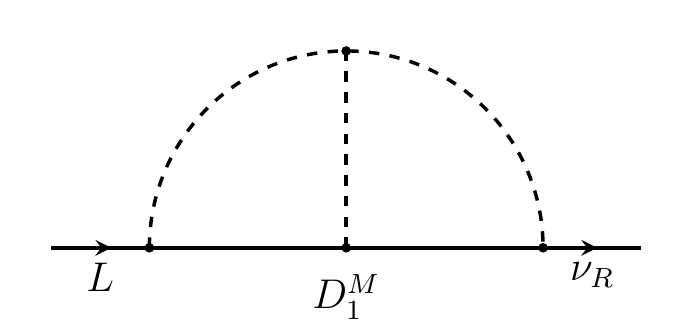}
    \includegraphics[scale=0.7]{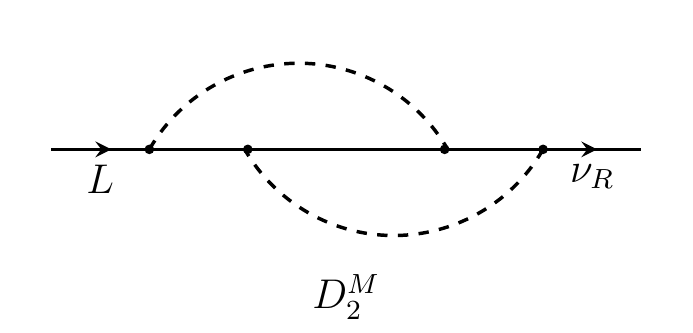}
    \includegraphics[scale=0.7]{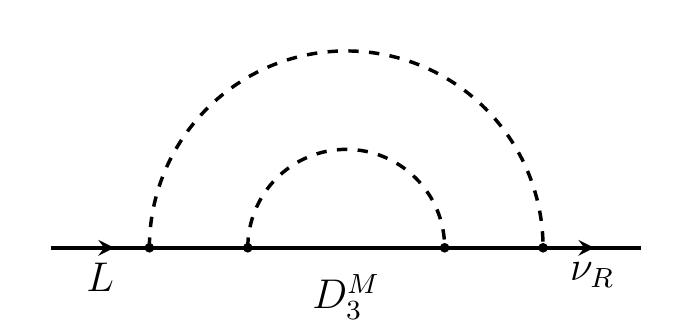}
    \\
    \includegraphics[scale=0.7]{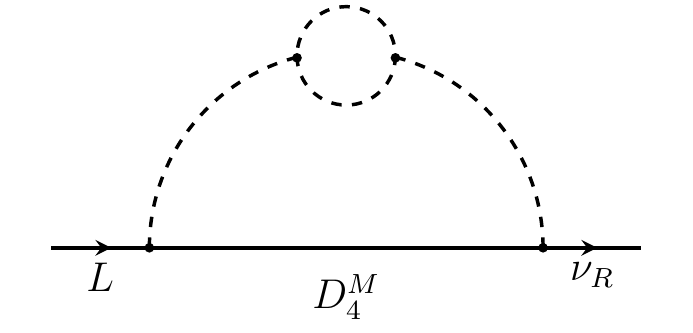}
    \includegraphics[scale=0.7]{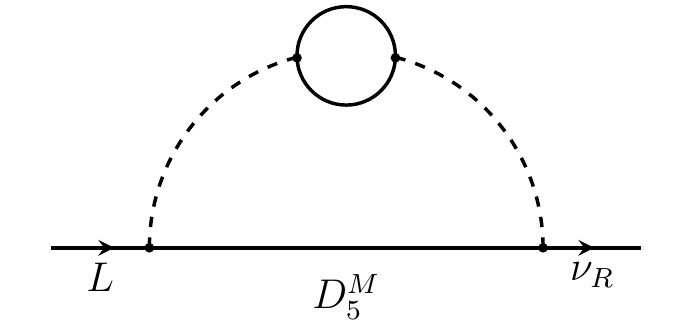}
    \includegraphics[scale=0.7]{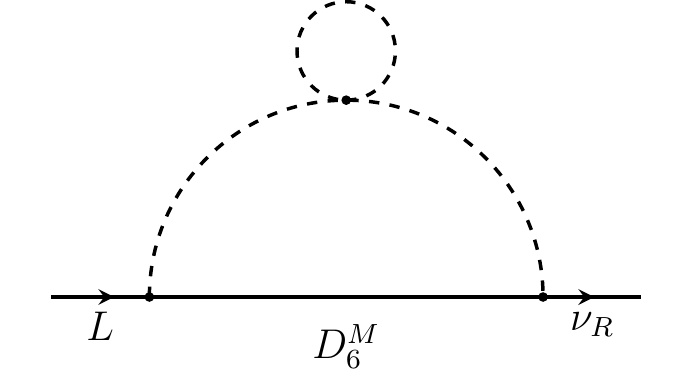}
    
    \caption{\footnotesize List of diagrams in the mass basis. Note that after removing the external Higgs line, there is no one-to-one correspondence between the diagrams in the gauge basis (Figs.~\ref{fig:diagrams 1},~\ref{fig:diagrams 2}~and~\ref{fig:diagrams 3}) and the mass diagrams given here.}
    \label{fig:massdiagrams}
\end{figure}

Finally, after spontaneous symmetry breaking the Higgs gets a vev generating the neutrino masses. Thus, the external scalar denoting the Higgs insertion is removed from the diagrams in the mass basis. The initial set of 18 genuine diagrams obtained in the electroweak basis is reduced to 6 diagrams. In figure~\ref{fig:massdiagrams} we show the list of 6 genuine mass diagrams.

All the mass diagrams can be computed analytically. Following the results of \cite{vanderBij:1983bw}, one can easily decompose any two-loop integral in figure \ref{fig:massdiagrams} in terms of just two master integral. A more detailed discussion is given in appendix \ref{sec:appendix 2loop}.

%% file: Section3_GeneratingModels.tex
%%%%%%%%%%%%%%%%%%%%%%%%%%%%%%%%%%%%%%%%%%%%%%%%%%%%%%%%%%%%%%%%%%%%%%%%%%%%%%%%%%%%%%%
%%%%%%%%%%%%%%%%%%%%%%%%%%%%%%%%%%%%%%%%%%%%%%%%%%%%%%%%%%%%%%%%%%%%%%%%%%%%%%%%%%%%%%%
\section{Generating models} \label{sec:generating models}%%%%%%%%%%%%%%%%%%%%%%%%%%%%%%
%%%%%%%%%%%%%%%%%%%%%%%%%%%%%%%%%%%%%%%%%%%%%%%%%%%%%%%%%%%%%%%%%%%%%%%%%%%%%%%%%%%%%%%
%%%%%%%%%%%%%%%%%%%%%%%%%%%%%%%%%%%%%%%%%%%%%%%%%%%%%%%%%%%%%%%%%%%%%%%%%%%%%%%%%%%%%%%

In this section we will discuss how to assign quantum numbers to the internal fields of the loops to obtain the \textit{model-diagrams}. It should be noted that on top of the gauge group of the Standard Model, an extra symmetry is needed to forbid the tree-level Dirac mass term $\bar{L} \phi^c \nu_R$, as well as to protect the Dirac nature of neutrinos. This can always be achieved by just one symmetry, which can be a residual subgroup of the global $B-L$ symmetry of the \sm \cite{Bonilla:2018ynb}. For now, in this section, we will only consider the \sm quantum numbers. The issue of the extra symmetry and its charge assignment will be discussed in the next section.

\begin{figure}[h!]
\centering
    \includegraphics[width=0.3\textwidth]{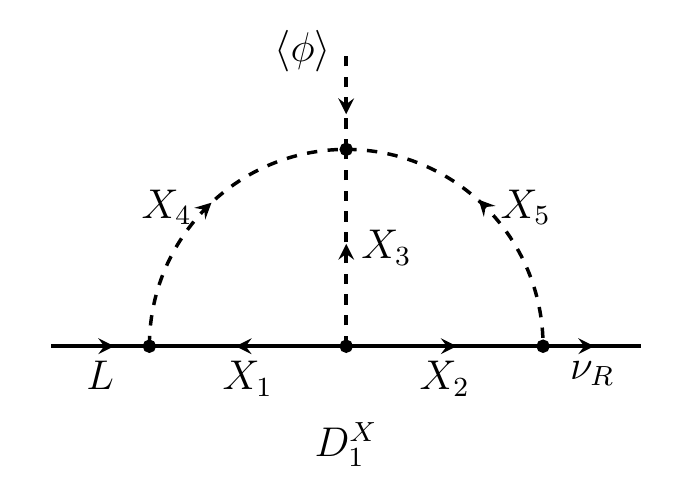}
    \includegraphics[width=0.3\textwidth]{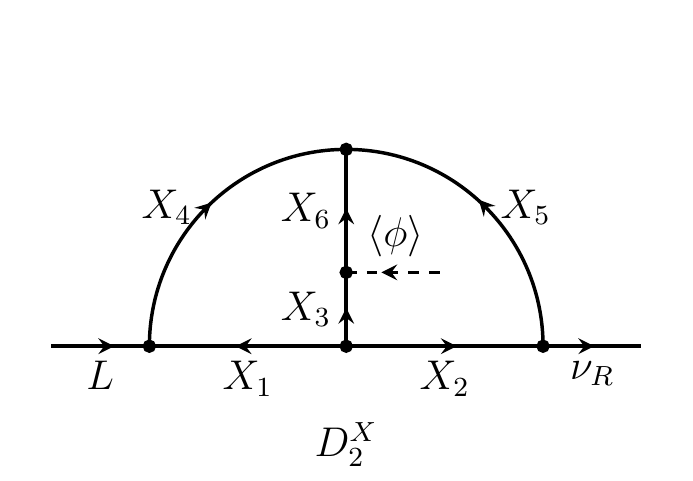}
    \includegraphics[width=0.3\textwidth]{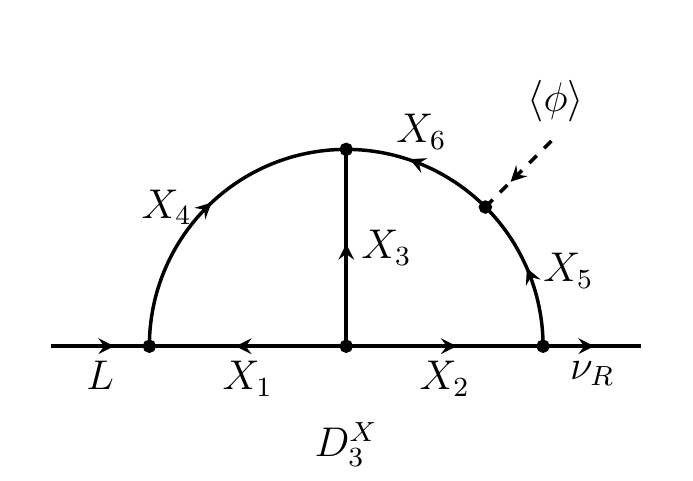}
    \vspace{-0.5cm}
    \\
    \includegraphics[width=0.3\textwidth]{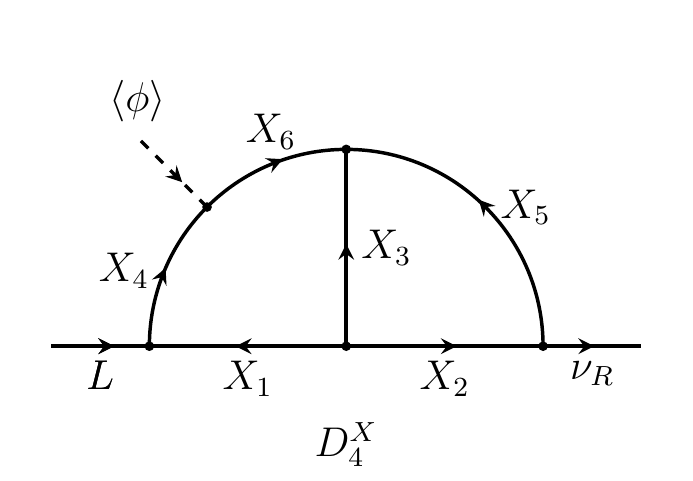}
    \includegraphics[width=0.3\textwidth]{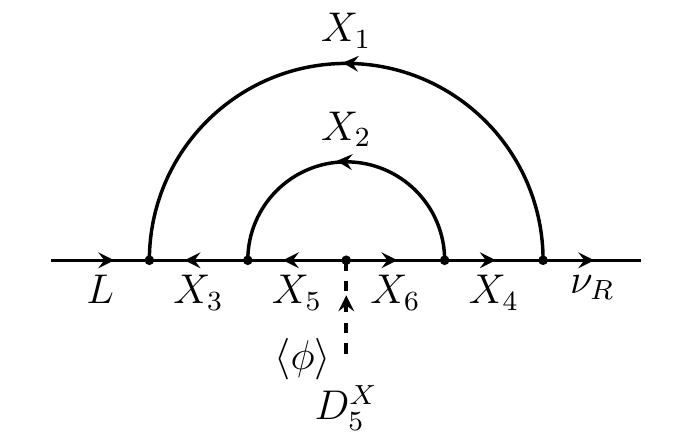}
    \caption{\footnotesize Auxiliary diagrams with a symbolic assignment of the internal fields at topology level. The external lines are assigned to $L$, $\phi$ or $\nu_R$, while the internal fields $X_i$ if represented by solid lines, can be either scalar or fermions. The arrows indicate the flowing of the quantum numbers.}
    \label{fig:diagrams-models-naming}
\end{figure}

\begin{figure}[h!]
\centering
    \includegraphics[width=0.7\textwidth]{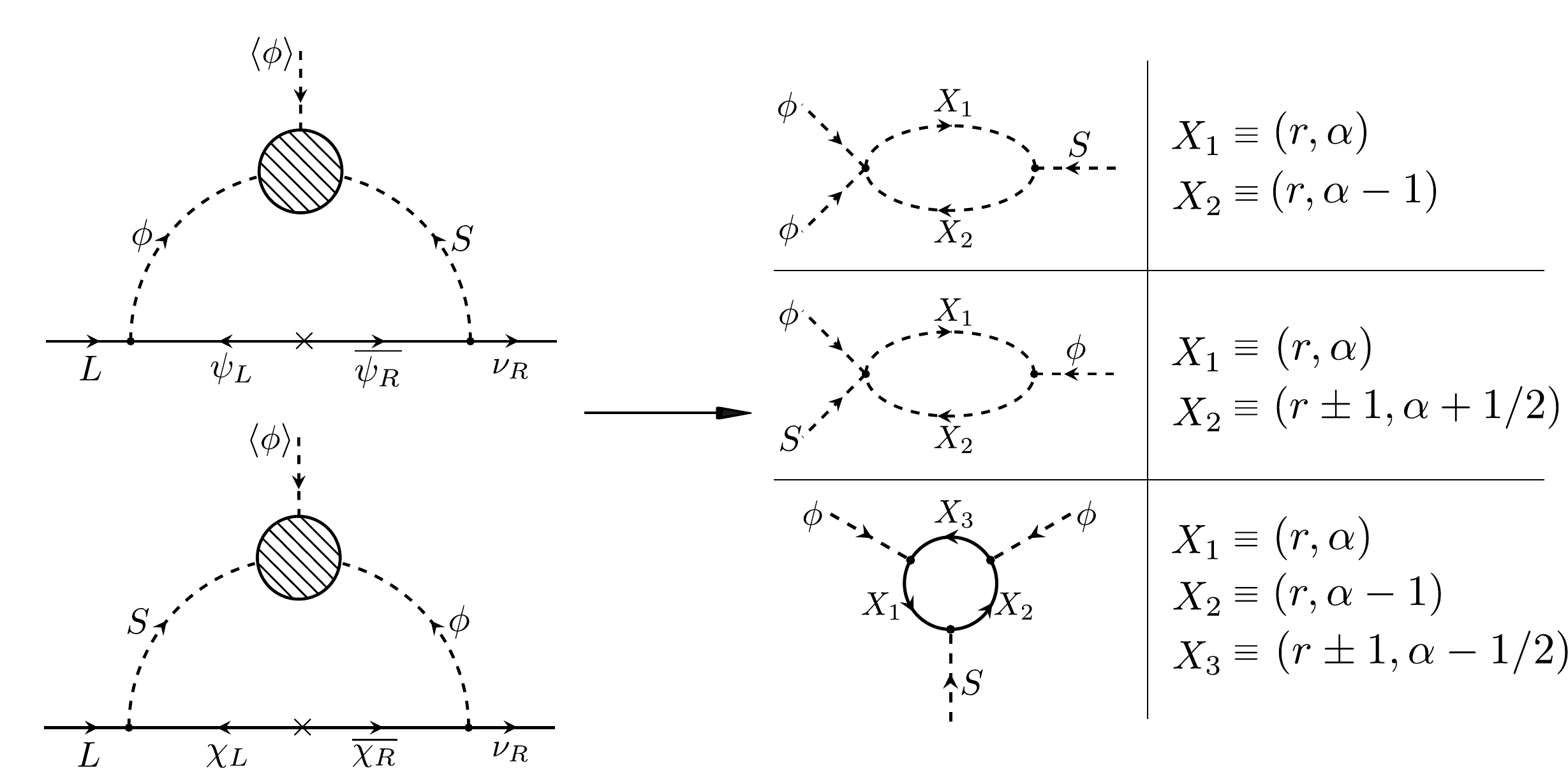}
    \caption{\footnotesize Auxiliary diagrams corresponding to those of figure~\ref{fig:diagrams 3}. The particle content depicted is the SM Higgs $\phi \equiv (\mathbf{2},1/2)$, $S\equiv(\mathbf{1},-1)$, $\psi_{L/R} \equiv (\mathbf{1},1)$, $\chi_{L/R} \equiv (\mathbf{2},-1/2)$ with charges under $SU(2)_L \times U(1)_Y$, while the unknown fields $X$ if solid lines, can be either scalars or fermions. We only consider colour singlets for simplicity. All the quantum numbers of the fields are determined once an input field $X_1 \equiv (r,\alpha)$ is given with $r>1$ (for $r=1$ only $r+1$ holds). See text for details.}
    \label{fig:models-loophole}
\end{figure}

Due to the large number of diagrams, it is more convenient to assign the quantum numbers at the topology level, while fixing the external fields, i.e. $L$, $\phi$ and $\nu_R$. This leaves us with the seven diagrams given in Figs.~\ref{fig:diagrams-models-naming} and~\ref{fig:models-loophole}. The separation into two distinct sets is done because the latter always requires the field $S\equiv(\mathbf{1},\mathbf{1},-1)$ in order to be genuine, which considerably constraints the possible fields running in the loop. See section~\ref{subsec:class 3} for details. 

In figure~\ref{fig:diagrams-models-naming}, the internal fields $X_i$ depicted as solid lines, can be either scalars or fermions. The only exception is in the $D_1^X$ diagram where, due to the quartic coupling, $X_3$, $X_4$ and $X_5$ can only be scalars and as such are drawn with dashed lines. This diagram corresponds directly to diagram T1-i. In the rest of the cases the correspondence with the diagrams in Figs.~\ref{fig:diagrams 1}~and~\ref{fig:diagrams 2} depends whether the fields $X_i$ are scalars or fermions. For example, $D^X_2$ corresponds to diagrams T2-i or T2-ii depending if $X_2$, $X_3$, $X_5$ and $X_6$ are scalars or fermions. $D_3^X$ corresponds to diagram T3-i, T3-iii, T3-vi and T3-ix; $D_4^X$ to T3-ii, T3-iv, T3-vii and T3-x; while diagrams T3-v and T3-viii are generated from $D_5^X$.

Since in all diagrams there are two-loops, there are two independent sets of quantum numbers (colour, $SU(2)_L$ representation and hypercharge), that need to be chosen in order to determine the rest of the fields. 
To fix the particle content, we start by assigning quantum numbers to $X_1$ and $X_2$. Once the gauge charges of these two fields are chosen, all the hypercharges of all other fields are automatically fixed. For the $SU(2)_L$ and $SU(3)_C$ representations, though, no general straightforward relation can be found since a product of two fields contains several irreducible representations. In spite of this, once the representations of $X_1$ and $X_2$ are chosen, the freedom of the other fields get severely restricted.
For simplicity we will work with colour singlets, because colour assignments can be trivially added taking into account that external fields are colour blind. We will explicitly omit this quantum number for the internal fields $X_i$.
As a side remark, note that every new fermion should have its corresponding vector-like partner to provide mass to them, as a fourth chiral family is excluded by Higgs production measurements and direct searches.

In table~\ref{tab:QN 1}, we give all possible fields assignments for a general hypercharge and up to $SU(2)_L$ triplets for the diagram $D_1^X$ of figure~\ref{fig:diagrams-models-naming}. For the rest of the diagrams $D^X_i$, the corresponding tables are given in Appendix~\ref{sec:appendix qn}.

\begin{table}[h!]
\centering
    \begin{tabular}{|*{5}{c|}}
        \hline\hline
        \multicolumn{5}{c}{\textbf{Hypercharge for $D_1^X$}} \\
        \hline
        \makebox[3em]{$X_1$} & \makebox[3em]{$X_2$} & $X_3$ & \makebox[3em]{$X_4$} & \makebox[3em]{$X_5$} \\
        \hline
        $\alpha_1$ & $\alpha_2$ & $-\alpha_1-\alpha_2$ & $\alpha_1-1/2$ & $\alpha_2$ \\
        \hline
    \end{tabular}
 \\[3ex]
    \begin{tabular}{|c||*{3}{c|}|*{3}{c|}|*{3}{c|}}
        \hline\hline
        \multicolumn{10}{c}{\textbf{$SU(2)_L$ representations for $D_1^X$}} \\
        \hline
        \backslashbox[3em]{$X_2$}{$X_1$} & \multicolumn{3}{c||}{1} & \multicolumn{3}{c||}{2} & \multicolumn{3}{c|}{3} \\
        \hline\hline
        & \makebox[2em]{$X_3$} & \makebox[2em]{$X_4$} & \makebox[2em]{$X_5$} & \makebox[2em]{$X_3$} & \makebox[2em]{$X_4$} & \makebox[2em]{$X_5$} & \makebox[2em]{$X_3$} & \makebox[2em]{$X_4$} & \makebox[2em]{$X_5$} \\
        \hline\hline
        1 & 1 & 2 & 1 & 2 & \slashbox[2.4em]{1}{3} & 1 & 3 & 2 & 1 \\
        \hline
        2 & 2 & 2 & 2 & \slashbox[2.4em]{1}{3} & \slashbox[2.4em]{1}{3} & 2 & 2 & 2 & 2 \\
        \hline
        3 & 3 & 2 & 3 & 2 & \slashbox[2.4em]{1}{3} & 3 & \slashbox[2.4em]{1}{3} & 2 & 3 \\
        \hline
    \end{tabular}
\caption{\footnotesize The $SU(2)_L$ and $U(1)_Y$ quantum numbers for the diagram $D_1^X$ (T1-i) of figure~\ref{fig:diagrams-models-naming}. Fixing the charges of the two fields $X_1$ and $X_2$, fixes the possible charges of all the other $X_{3-5}$ fields.  The possible $SU(2)_L$ representations (up to triplets) of the fields $X_1$ and $X_2$ are given in the first row and column of the second table. Their hypercharges are denoted by $\alpha_1$ and $\alpha_2$, respectively, in the first table. For the rest of the fields $X_{3-5}$ we give all the possible hypercharges and $SU(2)_L$ representations (up to triplets). For simplicity, all the fields are colour singlets.}
\label{tab:QN 1}
\end{table}

Table~\ref{tab:QN 1} is divided in two panels: hypercharge and $SU(2)_L$ representation. Hypercharges can be given in general by solving the system of equations for each vertex in terms of two input values $\alpha_1$ and $\alpha_2$, which are the hypercharges of $X_1$ and $X_2$ respectively, as shown in the upper panel of table~\ref{tab:QN 1}. In the lower panel, we show all the possible $SU(2)_L$ representations for the internal fields of $D_1^X$ up to triplets for different values of the quantum numbers of $X_1$ and $X_2$ in the first row and column, respectively. In the cases when several representations are possible (for example, $2 \otimes 2 = 1 \oplus 3$), the cell is subdivided to indicate that any of the two representations can be chosen.

Note that certain particular choices of the fields can generate lower order masses, i.e. tree or one-loop neutrino masses. In contrast to the Majorana case, here an additional model dependent symmetry is needed such that it forbids the lower order contributions. A judicious choice of the transformation of the fields under this symmetry and its appropriate breaking pattern is sufficient to ensure the genuineness of any 2-loop model generated from figure~\ref{fig:diagrams-models-naming} (see section~\ref{sec:classification} for details).

The diagrams that require the scalar field $S\equiv(\mathbf{1},\mathbf{1},-1)$ in order to be genuine, are shown in figure~\ref{fig:models-loophole}. We show all possible fields, along with their $SU(2)_L \times U(1)_Y$ charges, that close the diagrams in section~\ref{subsec:class 3}. Like before, we take all the fields to be colour singlets. 
As already explained, the main difference from the diagrams of the previous class (figure~\ref{fig:diagrams-models-naming}), is the necessity of a Higgs and the scalar $S$ running in the loop. 
In these diagrams, there is only one free set of quantum numbers, i.e. the quantum numbers of one of the fields running in the loop that generates the effective vertex $\phi\phi S$. As the external fields are fixed to be one $SU(2)_L$ singlet and two doublets, the quantum numbers of all the fields in the loop can be determined in general, once we pick the quantum numbers of any one of the remaining fields.  For example, choosing the $SU(2)_L \times U(1)_Y$ charges of $X_1$ field in figure~\ref{fig:models-loophole} as $X_1 \equiv (r,\alpha)$, $r$ denoting the $SU(2)_L$ representation and $\alpha$ the hypercharge, automatically fixes the 
possible charges of the remaining fields.
Note that unlike the previous class of models, here the coloured particles can only run in the small loop, see figure~\ref{fig:models-loophole} (right).  All the internal fields in this loop need to have the same $SU(3)_C$ representation since all the external fields are colour sinlgets.

As stated earlier, the set of models following from the topologies of figure~\ref{fig:models-loophole} are phenomenologically constrained. In all of them, one of the vector-like internal fermions must always have the \SM quantum numbers similar to either the quantum numbers of $L$ or $e^c$, i.e. $\chi \equiv (\mathbf{1},\mathbf{2},-1/2)$ and $\psi \equiv (\mathbf{1},\mathbf{1},1)$, respectively. Consequently, limits on their masses can be derived from collider searches \cite{Thomas:1998wy, Kumar:2015tna} and lepton flavour violating processes \cite{Falkowski:2013jya}, forcing their mass to be at least a  few TeV. Nevertheless, these constraints do not run in conflict with neutrino masses, since even for $\mathcal{O}(1)$ values of the couplings and the internal fermion masses in TeV range, the neutrino masses can easily be  $\mathcal{O}(0.1)$ eV scale \cite{Sierra:2014rxa}, as required by the current data.

%% file: Section4_Models.tex
%%%%%%%%%%%%%%%%%%%%%%%%%%%%%%%%%%%%%%%%%%%%%%%%%%%%%%%%%%%%%%%%%%%%%%%%%%%%%%%%%%%%%%%%%%%%%%%%%%%%%%%%
%%%%%%%%%%%%%%%%%%%%%%%%%%%%%%%%%%%%%%%%%%%%%%%%%%%%%%%%%%%%%%%%%%%%%%%%%%%%%%%%%%%%%%%%%%%%%%%%%%%%%%%%
\section{Example models} \label{sec:models}%%%%%%%%%%%%%%%%%%%%%%%%%%%%%%%%%%%%%%%%%%%%%%%%%%%%%%%%%%%%%
%%%%%%%%%%%%%%%%%%%%%%%%%%%%%%%%%%%%%%%%%%%%%%%%%%%%%%%%%%%%%%%%%%%%%%%%%%%%%%%%%%%%%%%%%%%%%%%%%%%%%%%%
%%%%%%%%%%%%%%%%%%%%%%%%%%%%%%%%%%%%%%%%%%%%%%%%%%%%%%%%%%%%%%%%%%%%%%%%%%%%%%%%%%%%%%%%%%%%%%%%%%%%%%%%

In this section we construct two example models to show in action the ideas discussed before. We have already presented the basic features and gauge charge requirements for the internal particles in the two-loop models. However, so far we have not explicitly discussed the role and nature of the symmetry or symmetries forbidding the tree-level coupling and/or protecting the Dirac nature of neutrinos. Since, as mentioned before, there are various options for such symmetries, a completely model independent approach is not possible. Let us now finally address the role of these symmetries by means of some example models.
  
There are many ways to arrange the additional symmetries of the model in such a way that all the necessary features are satisfied, namely neutrinos are Dirac particles and the leading contribution to its mass comes from the two-loop level. Another interesting feature which has been noticed before \cite{Chulia:2016ngi, Chulia:2016giq, Bonilla:2018ynb, Bonilla:2019hfb} and we will explicitly discuss is section~\ref{sec:dm} is the connection between Dirac nature of neutrinos and dark matter stability. 
If chosen correctly, the symmetry protecting the Diracness of neutrinos can also forbid the decay of the dark matter, ensuring its stability. Thus, the additional symmetry can play multiple roles. Furthermore, as has been discussed in \cite{Bonilla:2018ynb}, this symmetry can also forbid the lower order mass terms. Additionally, it need not be a brand new symmetry and can just be a residual subgroup of the global $U(1)_{B-L}$ symmetry already present in the Standard Model. We will discuss the Diracness-dark matter stability connection in more details in section~\ref{sec:dm}.

The two  examples we show in this section employ the discrete abelian cyclic $\mathcal{Z}_4$ group as the symmetry protecting the Dirac nature of neutrinos. Although not necessary, this  symmetry can be a residual subgroup of the $U(1)_{B-L}$ symmetry of the \sm \cite{Chulia:2016ngi, Chulia:2016giq, CentellesChulia:2017koy, Dasgupta:2019rmf} or of some other $U(1)_X$ symmetry \cite{Ma:2019yfo}. The choice of the $\mathcal{Z}_4$ symmetry is done keeping in mind the Diracness connection to dark matter stability to be discussed in the next section.
It is worth to notice that if this symmetry is taken as $\mathcal{Z}_2$ then neutrinos will be Majorana fields \cite{Hirsch:2017col}. Taking  it to be $Z_3$ will necessarily lead either to decaying dark matter or to the existence of an accidental symmetry that stabilizes dark matter \cite{Bonilla:2018ynb,Bonilla:2019hfb}. Therefore $\mathcal{Z}_4$ is the smallest group that achieves simultaneously the stability of the dark matter while protecting the Dirac nature of neutrinos. 

For both models the lepton doublets $L_i $ and right-handed neutrinos $\nu_{R,i}$ transform as
``$\mathcal{Z}_4$ odd'' particles i.e. $z^1 = e^{i \pi/2}=i$ under the $\mathcal{Z}_4$ symmetry with $z^4=1$.\footnote{We call $\mathcal{Z}_4$ odd the fields that transform as odd powers, i.e. the fields transforming as $z^1 \equiv i$ or as $z^3 \equiv -i$ under the $\mathcal{Z}_4$ symmetry. Similarly, $\mathcal{Z}_4$ even are the fields transforming as even powers i.e. $z^0 \equiv 1$ or as $z^2 \equiv -1$ under the $\mathcal{Z}_4$ symmetry. } This automatically forbids all Majorana mass terms for the neutrinos at all dimensions and loop orders, and ensures they are Dirac particles. 
We also add a $\mathcal{Z}_2$ symmetry whose role is to forbid the tree-level mass term for neutrinos. This symmetry will be softly broken to allow for the two-loop realization of the operator $\overline{L} \phi^c \nu_R$ \cite{Chulia:2016ngi}. We would like to remark that this additional $\mathcal{Z}_2$ symmetry is not always necessary to forbid the tree-level mass term~\cite{Ma:2014qra, Ma:2015mjd,Bonilla:2018ynb,Bonilla:2019hfb}, however we have added it in order to keep the discussion simple. 
Further note that, as shown in \cite{Bonilla:2018ynb}, all these features can be obtained using only the $B-L$ symmetry without the need of extra symmetries. Although this construction is appealing because of its economic symmetry inventory, it is conceptually a bit more involved than the simple one we choose here as an example.

%%%%%%%%%%%%%%%%%%%%%%%%%%%%%%%%%%%%%%%%%%%%%%%%%%%%%%%%%%%%%%%%%%%%%%%%%%%%%%%%%%%%%%%%%%%%%%
\subsection{A genuine two-loop Dirac neutrino mass model}%%%%%%%%%%%%%%%%%%%%%%%%%%%%%%%%%%%%%
%%%%%%%%%%%%%%%%%%%%%%%%%%%%%%%%%%%%%%%%%%%%%%%%%%%%%%%%%%%%%%%%%%%%%%%%%%%%%%%%%%%%%%%%%%%%%%

From the diagrams given in section~\ref{subsec:class 1}, we choose T1-i in figure~\ref{fig:diagrams 1} to illustrate how a simple genuine model can be built. As described before, in contrast to the diagrams of sections~\ref{subsec:class 2} and \ref{subsec:class 3}, the main characteristic of the models generated from these diagrams is that, a priori, there is no restriction on the possible internal fields or the position of the soft symmetry breaking terms. One should only be careful about choosing the charges of the internal fermions in such a way that the leading contribution comes at the two-loop order.\footnote{For instance, avoid new fermions $F$ with quantum numbers that allow the vertex $\overline{L}\phi^c F$.}

\begin{figure}[h!]
\centering
    \includegraphics[scale=0.75]{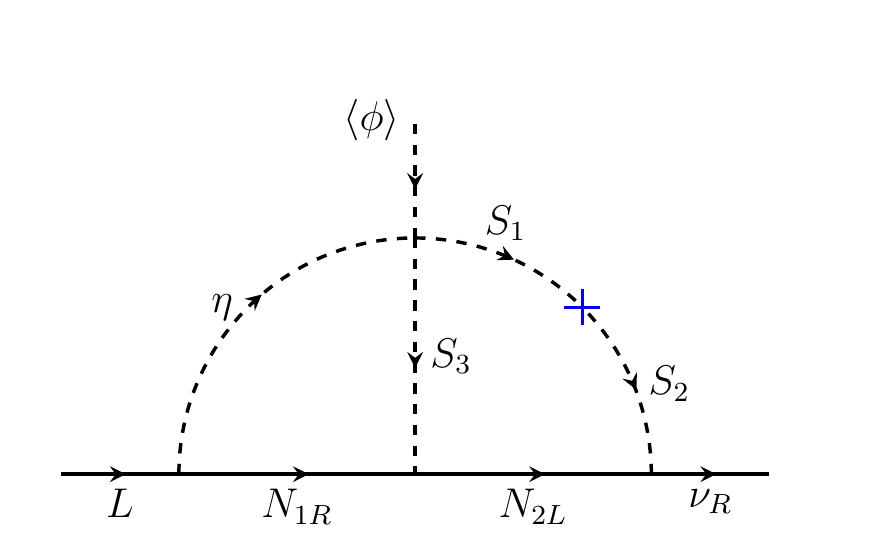}
    \caption{\footnotesize Completely genuine two-loop diagram that gives mass to neutrinos. The blue cross marks the soft breaking term of the $\mathcal{Z}_2$ symmetry that allows the loop realization of the operator $\overline{L} \phi^c \nu_R$ forbidding the tree-level.}
    \label{fig:softly} 
\end{figure}

Following table~\ref{tab:QN 1} for the simplest case when  $X_1$ and $X_2$ are $SU(2)_L$ singlets, we construct the model of figure~\ref{fig:softly}, whose  particle content and  relevant quantum numbers are given in table~\ref{tab:model genuine}. Two extra symmetries, apart from those of the Standard Model gauge group are added, a $\mathcal{Z}_4$ and a $\mathcal{Z}_2$. The former ensures the Dirac nature of neutrinos and as we will discuss in section~\ref{sec:dm}, at the same time it also provides the stability of dark matter, while the latter is related to the smallness of neutrino masses, forbidding the tree-level mass operator $\overline{L} \phi^c \nu_R$. The $\mathcal{Z}_2$ symmetry is softly broken in order to allow the loop realization of figure~\ref{fig:softly}. Including all the soft-breaking terms to the Lagrangian means that we have to add the mass term $S_2^\dagger S_1 + \text{h.c.}$, depicted as a blue cross on the diagram.\footnote{The soft term $\phi\eta^\dagger S_2$ should be added too for consistency, although it plays no role in the neutrino mass generation or the dark matter stability.}

\begin{table}
\begin{center}
\begin{tabular}{| c || c | c | c | c |}
  \hline
&   \hspace{0.1cm}  Fields     \hspace{0.1cm}       &  \hspace{0.4cm}  $SU(2)_L \times U(1)_Y$     \hspace{0.4cm}    & \hspace{0.4cm}   $\mathcal{Z}_4$            \hspace{0.4cm}   &  \hspace{0.4cm}  $\mathcal{Z}_2$            \hspace{0.4cm}                            \\
\hline \hline
\multirow{4}{*}{ \begin{turn}{90} Fermions \end{turn} } 
&   $L$        	     &   ($\mathbf{2}, {-1/2}$)      &   $i$ & $+$\\
&   $\nu_{R}$        &   ($\mathbf{1}, {0}$)         &   $i$ & $-$ \\
&   $e_{R}$          &   ($\mathbf{1}, {-1}$)        &   $i$ & $+$  \\
&   $E_{R}$         &   ($\mathbf{1}, {-1}$)        &   $1$ & $+$   \\
&   $E_{L}$    	 &   ($\mathbf{1}, {-1}$)        &   $1$ & $+$    \\
&   $N_{L}$      	 &   ($\mathbf{1}, {0}$)         &   $1$ & $+$     \\
\hline \hline                                                                               
\multirow{5}{*}{ \begin{turn}{90} Scalars \end{turn} } 
& $\phi$  	       	 &  ($\mathbf{2}, {1/2}$)        & $1$   & $+$        \\
& $S_1$          	 &  ($\mathbf{1}, {0}$)          & $i$   & $+$         \\
& $S_2$              &  ($\mathbf{1}, {0}$)          & $i$   & $-$          \\
& $S_3$              &  ($\mathbf{1}, {1}$)          & $1$   & $+$           \\
& $\eta$             &  ($\mathbf{2}, {1/2}$)        & $i$   & $+$            \\
    \hline
  \end{tabular}
\end{center}
\caption{\footnotesize Particle content of the completely genuine example model. The gauge charges along with the $\mathcal{Z}_4 \times \mathcal{Z}_2$ charges are also shown. All the fields listed in the table are $SU(3)_C$ singlets. The role of $\mathcal{Z}_4$ is to protect Diracness and to stabilize the dark matter candidate. The lightest  particle out of $S_1$, $S_2$, the neutral component of $\eta$ and the Majorana fermion $N_{L}$. The $\mathcal{Z}_2$ symmetry forbids the neutrino tree-level mass $\overline{L} \phi^c \nu_R$ and it is  softly broken to allow the two-loop realization of such operator.}
 \label{tab:model genuine}
\end{table}%

The $\mathcal{Z}_4$ charges are chosen to forbid the Majorana mass term for $\nu_R$. It also forbids the mixing of the internal fermions with the Standard Model fermions. This avoids undesirable  terms that may mix new fermions with the charged leptons or the neutrinos. Moreover, $\mathcal{Z}_4$ ensures the stability of the dark matter candidate, in our case the lightest of the ``$\mathcal{Z}_4$ odd'' scalars and ``$\mathcal{Z}_4$ even'' fermions i.e. lightest among ($S_1$, $S_2$, $\eta^0$, $N_{L}$), as we will show in section~\ref{sec:dm}.

The new fermions in the loop are of two types. $E$ is a massive, $SU(2)_L$ singlet vector-like fermion carrying hypercharge. Although its quantum numbers are the same as those of right-handed charged leptons, the $\mathcal{Z}_4$ symmetry forbids their mixing. Since it is electrically charged, it cannot be the dark matter candidate. Therefore, its mass has to be taken sufficiently high. The fermion $N_{L}$ is also a $SU(2)_L$ singlet fermion but carries no hypercharge. Owing to its quantum charges, one can write down a Majorana mass term for it and hence it's right-handed partner is not needed to give it mass. Being a neutral $\mathcal{Z}_4$ even fermion, it can be a good dark matter candidate.

The scalars running in the loop, $\eta$ and $S_i$, must have exactly zero vev in order to avoid breaking the $\mathcal{Z}_4$ symmetry and therefore losing all the attractive features associated to it. Moreover, given their charges under $\mathcal{Z}_4$, the lightest can be a good dark matter candidate, except $S_3$ which decays to the Standard Model.

Moving on, the neutrino mass of the diagram in figure~\ref{fig:softly} is generated from the following terms of the Lagrangian,
\begin{eqnarray} \label{eq:lagrangian 1}
\mathcal{L}_\nu & = & (Y_1)_{\alpha i}\, \overline{L}_\alpha E_{R_i} \eta 
+ (Y_2)_{ i \alpha}\, \overline{N}_{L_i} \nu_{R_\alpha}  S_2^\dagger 
+ (Y_{12})_{ij}\, \overline{N}_{L_i} E_{R_j} S_3 
+ \text{h.c.}
    \nonumber
    \\
&+& (\mathcal{M}_E)_{ij} \, \overline{E}_{L_i} E_{R_j} 
+ (\mathcal{M}_N)_{ij} \, \overline{N}^c_{L_i} N_{L_j} + \text{h.c.}
    \\
    \nonumber
    &+& \left[ \lambda \, \eta \phi S_1^\dagger S_3^\dagger + \mu_{12}^2 S_2^\dagger S_1  + \text{h.c.} \right] 
    + m_\eta^2 \eta^\dagger \eta + \sum\limits_{k=1}^3 m_{S_k}^2 S_k^\dagger S_k
    + ...\;,
\end{eqnarray}
with $\alpha=1,2,3$ and where the term $\mu_{12}$ breaks softly the $\mathcal{Z}_2$ symmetry. Other terms of the Lagrangian are not explicitly given, as they are not relevant for the neutrino mass generation. 
At this point, there is no need to fix the number of internal fermion copies. Nevertheless, given the fact that at least two neutrinos should have mass, the minimal choice in order to fit neutrino data would be $i,j=1,2$.
%For simplicity we will consider that $Y_{12} = Y_{12}'$. 
Consequently, the effective Yukawa is given by
\begin{equation} \label{eq:yukawa nu 1}
    \left( Y_\nu \right)_{\alpha\beta} \approx \frac{1}{(16\pi^2)^2}\, \lambda\, \frac{\mu_{12}^2}{m_{S_3}^2} \left[ \frac{M_{Ei} M_{Nj}}{m_{S_3}^2} F^{(1)}_{ij} + F^{(2)}_{ij} \right]\, (Y_1)_{\alpha i} (Y_{12})_{ij} (Y_2)_{\beta j} ,
\end{equation}
with $M_{Ei}$ and $M_{Ni}$ the mass eigenstates of the $i$-copy of the fermions $E$ and $N_L$, respectively. The dimensionless loop integrals $F_{ij}$ are obtained directly in the mass insertion approximation assigning momenta to the internal fields as,
\begin{subequations} \label{eq:F integrals 1}
    \begin{equation} %\label{eq:}
        F^{(1)}_{ij} = m_{S_3}^4 \iint\limits_{(k,q)} \frac{ 1 }{ (k^2-m_\eta^2) (k^2-M_{Ei}^2) (q^2-M_{Nj}^2) (q^2-m_{S_1}^2) (q^2-m_{S_2}^2) ((k+q)^2-m_{S_3}^2) },
    \end{equation}
    \begin{equation} %\label{eq:}
        F^{(2)}_{ij} = m_{S_3}^2 \iint\limits_{(k,q)} \frac{ k \cdot q }{ (k^2-m_\eta^2) (k^2-M_{Ei}^2) (q^2-M_{Nj}^2) (q^2-m_{S_1}^2) (q^2-m_{S_2}^2) ((k+q)^2-m_{S_3}^2) },
    \end{equation}
\end{subequations}
with the shorthand $\int\limits_k \equiv (16\pi^2)\int d^4\!k / (2\pi)^4$. Both integrals can be written in terms of simple one-loop and two-loop integrals, for which analytical expressions can be found. The decomposition of the integrals \eqref{eq:F integrals 1} is done as an example of how to compute two-loop radiative masses in the Appendix~\ref{sec:appendix 2loop}.

In order to fit the neutrino oscillation data \cite{deSalas:2017kay}, we need at least two copies of $E$ and $N_L$, so that $Y_\nu$ is a rank-2 matrix, giving masses to two neutrinos. We have then three rank-2 matrices ($Y_1$, $Y_2$, $Y_{12}$) with enough freedom to fit the two neutrino mass squared differences, along with the three mixing angles and phases. However, here we will only consider the case 
with one massive neutrino, by assuming no hierarchy or flavour structure in the Yukawas and just one copy of the new fermions, i.e. $Y_1 = Y_2 = Y_{12} = Y$. This is done in order to simplify our analysis and show the characteristic neutrino mass scale $m_\nu$ in this type of models. 

\begin{figure}
\centering
    \includegraphics[width=0.5\textwidth]{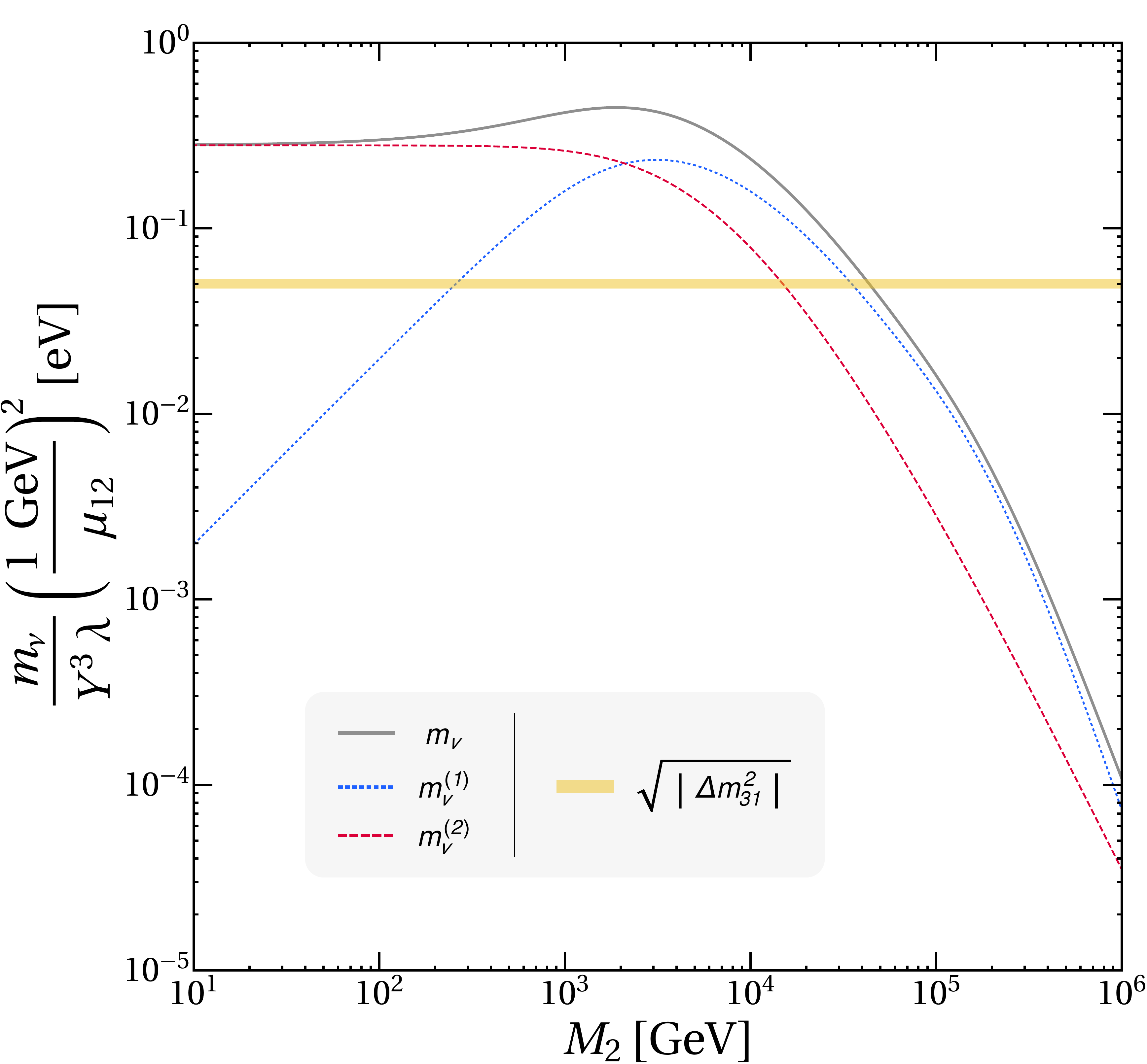}
    \caption{\footnotesize Neutrino mass scale $m_\nu$ (solid line) for the diagram of figure~\ref{fig:softly} with respect to the mass of $N_L$. The rest of the masses are $\mathcal{O}(1)$ TeV. The contributions coming from both $F^{(a)}$, 
    see~\eqref{eq:yukawa nu 1}, has been separated (dashed lines). The atmospheric mass scale (yellow line) is plotted for comparison. See text for details.}
    \label{fig:NuMass plot 1} 
\end{figure}

The behaviour of the neutrino mass is given in figure~\ref{fig:NuMass plot 1} in terms of the couplings of the model and the mass of $N_{L}$ field. Here, we consider that the masses of the rest of the internal fields are of order $1$ TeV. The atmospheric scale $\sqrt{|\Delta m_{13}^2 |} \approx 0.05$ eV is plotted for comparison. The dashed lines $m_\nu^{(a)}$ represent the neutrino mass scale when only the loop integral $F^{(a)}$ is considered, while the solid line is for the complete mass equation \eqref{eq:yukawa nu 1}.
In figure~\ref{fig:NuMass plot 1} we see the distinct behaviour of $F^{(1)}$ and 
$F^{(2)}$ due to the different numerators. Also, notice how the function decreases when $M_N$ overtake the rest of the masses of $\mathcal{O}(1)$ TeV and its propagator starts dominating the integral. 
If all couplings are taken to be  $\mathcal{O}(1)$, the mass scale should be about $100$ TeV. Nevertheless, the cubic dependence of the neutrino mass with the Yukawas can lose this scale considerably, allowing masses of order $1$ TeV or below accessible at colliders. 

%%%%%%%%%%%%%%%%%%%%%%%%%%%%%%%%%%%%%%%%%%%%%%%%%%%%%%%%%%%%%%%%%%%%%%%%%%%%%%%%%%%%%
\subsection{Model exploiting the non-local realization of $\phi\phi S$}%%%%%%%%%%%%%%%%%%%%%
%%%%%%%%%%%%%%%%%%%%%%%%%%%%%%%%%%%%%%%%%%%%%%%%%%%%%%%%%%%%%%%%%%%%%%%%%%%%%%%%%%%%%

Now we move to a different class of diagrams, those depicted in figure~\ref{fig:diagrams 2}. Models generated from these diagrams need certain fields in order to be genuine. As explained in section~\ref{subsec:class 3}, they contain a one-loop three scalar vertex with one external Higgs. Such loops are reducible unless the other scalars are another Higgs and the charged singlet $S\equiv(\mathbf{1},\mathbf{1},-1)$, realizing the loop effective coupling $\phi\phi S$ (see figure~\ref{fig: example HHS}).

As a simple example of how these models work, we will take the diagram T3-xi in figure~\ref{fig:diagrams 3} and add to the Standard Model the particle content given in table~\ref{model HHS}. The role of the cyclic $\mathcal{Z}_4$ and $\mathcal{Z}_2$ symmetries is analogous to the previous model.

\begin{table}[h!]
\begin{center}
\begin{tabular}{| c || c | c | c | c |}
  \hline 
&   \hspace{0.1cm}  Fields     \hspace{0.1cm}       &  \hspace{0.4cm}  $SU(2)_L \times U(1)_Y$     \hspace{0.4cm}    & \hspace{0.4cm}   $\mathcal{Z}_4$            \hspace{0.4cm}   &  \hspace{0.4cm}  $\mathcal{Z}_2$            \hspace{0.4cm}                            \\
\hline \hline
\multirow{4}{*}{ \begin{turn}{90} Fermions \end{turn} }
&   $L$        	     &   ($\mathbf{2}, {-1/2}$)      &   $i$ & $+$ \\	
&   $\nu_{R}$        &   ($\mathbf{1}, {0}$)         &   $i$ & $-$  \\
&   $e_{R}$          &   ($\mathbf{1}, {-1}$)        &   $i$ & $+$   \\
&   $E_{R}$          &   ($\mathbf{1}, {-1}$)        &   $i$ & $+$    \\
&   $E_{L}$    	     &   ($\mathbf{1}, {-1}$)        &   $i$ & $+$     \\
\hline \hline                                                                              
\multirow{5}{*}{ \begin{turn}{90} Scalars \end{turn} } 
& $\phi$  	       	 &  ($\mathbf{2}, {1/2}$)        & $1$   & $+$        \\		
& $S_1$          	 &  ($\mathbf{1}, {1}$)          & $1$   & $-$         \\
& $S_0$              &  ($\mathbf{1}, {0}$)          & $i$   & $+$          \\	
& $S_1'$             &  ($\mathbf{1}, {1}$)          & $i$   & $+$           \\	
& $\eta$             &  ($\mathbf{2}, {1/2}$)        & $i$   & $+$            \\
    \hline
  \end{tabular}
\end{center}
\caption{\footnotesize Particle content of the example $\phi\phi S$ model. The gauge charges along with the $\mathcal{Z}_4 \times \mathcal{Z}_2$ charges are also shown. All the fields listed in the table are $SU(3)_C$ singlets. Again, the role of $\mathcal{Z}_4$ is to protect Diracness and to stabilize the dark matter candidate: the lightest out of $S_0$ and the neutral component of $\eta$. The $\mathcal{Z}_2$ symmetry forbids the neutrino tree-level mass $\overline{L} \phi ^c \nu_R$ and it is  softly broken to allow the two-loop realization of such operator.}
 \label{model HHS}
\end{table}%

Note that the new fermions have the same gauge charges as the right-handed charged leptons and, therefore, they mix. This mixing has to be controlled in order for the model to be phenomenologically viable. Looking to the relevant Lagrangian terms,
\begin{equation}
 \mathcal{L}_{\nu N} \, = \, (Y_e)_{\alpha\beta} \, \overline{L}_\alpha \phi \,  e_{R\beta} + (Y_1)_{\alpha i}  \, \overline{L}_\alpha \phi E_{Ri} + (\mathcal{M}_E)_{ij} \, \overline{E}_{Li} E_{Rj} + \mathcal{X}_{\alpha i} \, \overline{E}_{Li} \, e_{R\alpha} + \text{h.c.},
\end{equation}
which can be written in matrix form as
\begin{equation}
    \left (\begin{matrix} \overline{L} & \overline{E}_L \end{matrix} \right)
                                    \left (\begin{matrix}
                                    Y_e v & Y_1 v \\
                                    \mathcal{X} & \mathcal{M}_E
                                   \end{matrix} \right)
                                   \left (\begin{matrix}
                                   e_R \\
                                   E_R
                                 \end{matrix} \right).
\end{equation}
Here, $\alpha, \beta = 1,2,3$ and $(i,j)$ the number of copies of $E$. Taking the Yukawa matrices of order $1$, it is easy to see that if the elements of matrices $\mathcal{X}$ and $\mathcal{M}_E$ are bigger than the \sm vev $v$, then the mixing in the left-handed sector will be sufficiently small to avoid collider constraints. The phenomenology of vector-like singlet leptons has been extensively studied in the literature, setting limits to their masses around $\sim 100$ GeV \cite{Kumar:2015tna, Falkowski:2013jya}. Therefore it is safe to consider the mass of the charged fermions to be around or above the TeV scale. Lepton flavour violating processes can set also stringent limits on the mass depending on the value of $Y_1$. Nevertheless, given our model, lepton flavour violation can be hidden, as there are enough free parameters between $Y_1$ and $Y_2$ in order to fit neutrino data while suppressing significantly any flavour violating signal. A more thorough study, beyond the scope of this work, would be needed to give a reasonable limit.

\begin{figure}
\centering
    \includegraphics[width=0.45\textwidth]{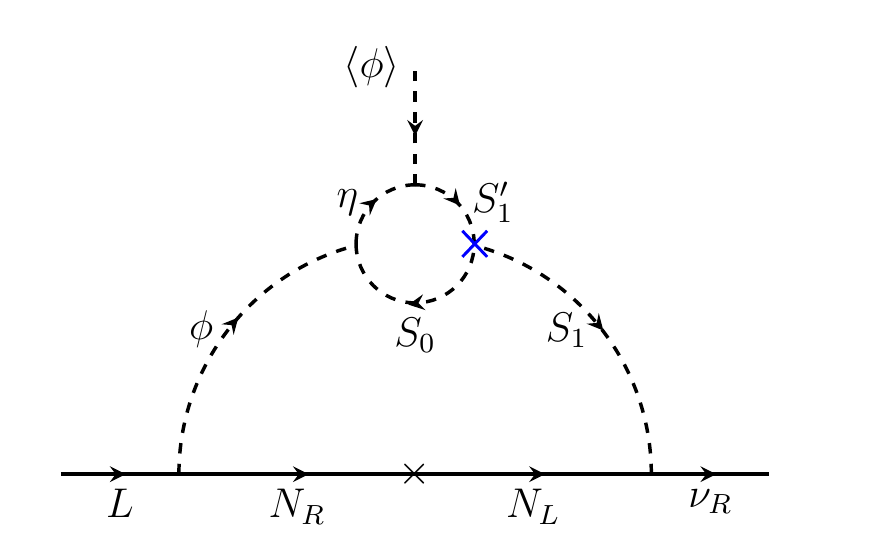}
    \includegraphics[width=0.45\textwidth]{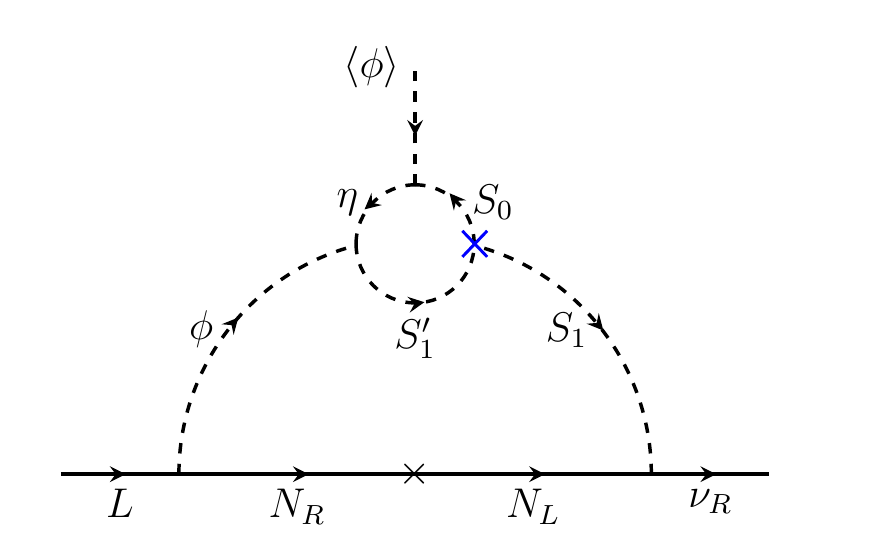}
    \caption{\footnotesize Leading contributions to neutrino masses. The blue cross marks the soft breaking term of the $\mathcal{Z}_2$ symmetry that allows the two-loop realization of the operator $\overline{L} \phi^c \nu_R$. Note that the small scalar loop cannot be reduced into a tree-level vertex.}
    \label{fig:loopHHS}
\end{figure}

In the neutrino sector, as explained before, the tree-level mass term $\overline{L} \phi ^c \nu_R$ is forbidden by the $\mathcal{Z}_2$ symmetry. Indeed, this symmetry will forbid all the loop realizations of such operators unless it is softly broken. 
Once we allow the soft breaking of $\mathcal{Z}_2$, we have to add only one extra term to the Lagrangian, $S_1^\dagger S_1' S_0^\dagger$. This term is essential and leads to the neutrino mass diagrams depicted in figure~\ref{fig:loopHHS}.

One could be tempted to add also the tree-level coupling $\phi \phi S_1$, which is allowed by the gauge symmetry and $\mathcal{Z}_4$ and breaks $\mathcal{Z}_2$ only softly. However, this term vanishes because the contraction $\phi\phi$ going to a singlet is completely antisymmetric. Therefore the leading contribution to neutrino mass will be the two-loop diagrams shown in figure~\ref{fig:loopHHS} (see section~\ref{subsec:class 3} for details).

As we will argue in section~\ref{sec:dm}, the lightest among the neutral component of $\eta$, $S1'$ and $S_0$ will be stable and thus a good dark matter candidate. In this model all the dark matter candidates are scalars.

\begin{figure}
\centering
    \includegraphics[width=0.5\textwidth]{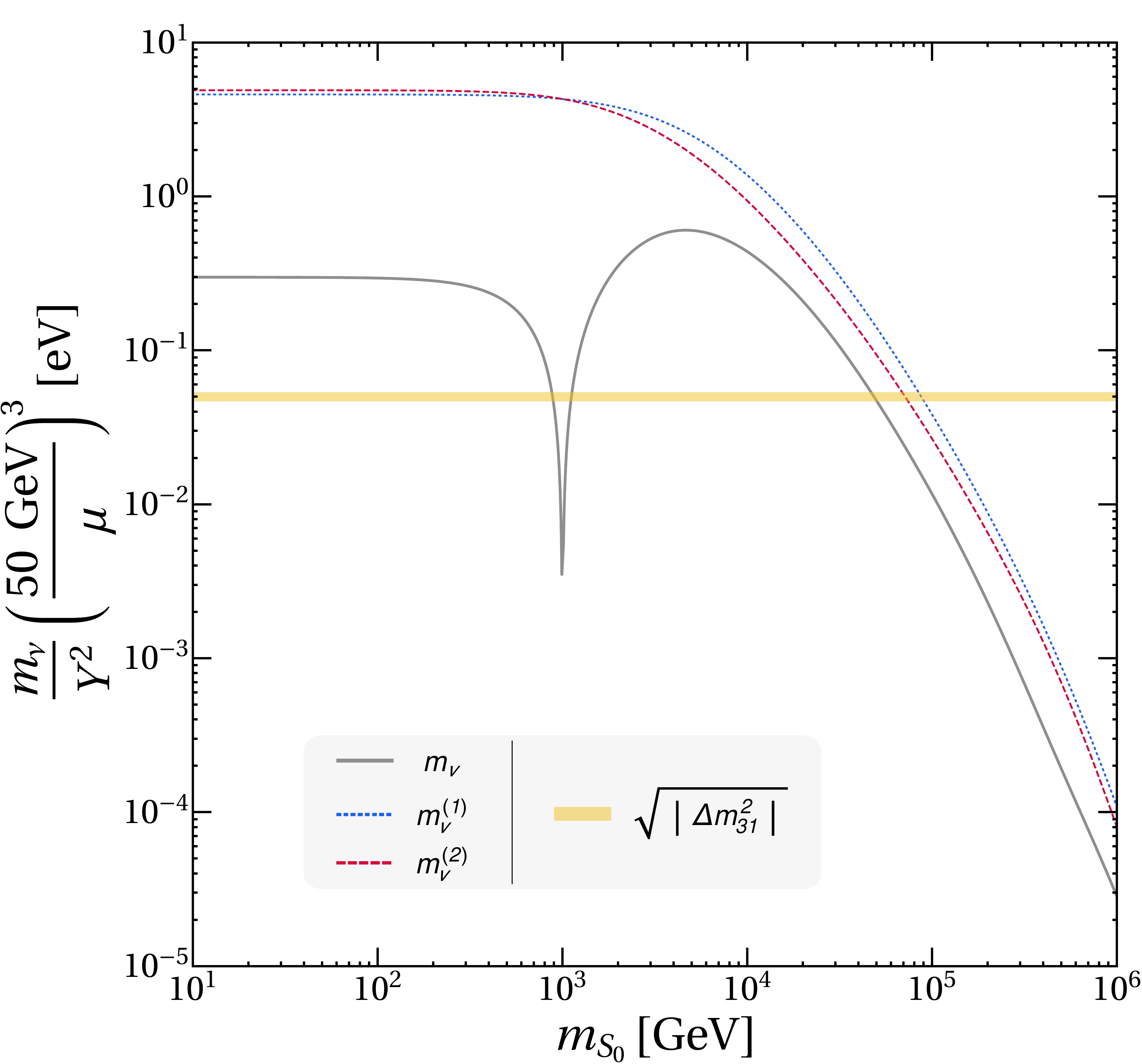}
    \caption{\footnotesize Neutrino mass scale $m_\nu$ (solid line) for the diagrams of figure~\ref{fig:loopHHS} with respect to the mass of the neutral scalar $S_0$. The rest of the masses are $\mathcal{O}(1)$ TeV. Both contributions are separated and represented as dashed lines. The atmospheric mass scale (yellow line) is plotted for comparison. See text for details.}
    \label{fig:NuMass plot 2} 
\end{figure}

Regarding neutrino masses. The main feature of these class of models is that given the loop vertex with two identical Higgs $\phi$, there are always two contributions simply interchanging both Higgs. Moreover, both contributions have a relative minus sign due to the antisymmetric nature of $SU(2)_L$. Precisely, the minus sign comes from the coupling $\phi \eta S_1^{'\dagger} \equiv \epsilon^{ij} \phi_i \eta_j S_1^{'\dagger} = (\phi^+ \eta^0 - \phi^0 \eta^+) S_1^{'\dagger}$. This can produce a cancellation between both diagrams, leading to a suppression of the neutrino mass scale as can be seen in figure~\ref{fig:NuMass plot 2}.

The corresponding terms of the Lagrangian that appear in the diagrams of figure~\ref{fig:loopHHS} are 
\begin{eqnarray} \label{eq:lagrangian 2}
    \mathcal{L} &=& (Y_1)_{\alpha i}\, L_\alpha \overline{E}_{Ri} \phi^\dagger + (Y_2)_{\alpha i}\, \overline{\nu_R}_\alpha E_{Li} S_1 + (\mathcal{M}_E)_{ij} \, \overline{E}_{Ri} E_{Lj} + \text{h.c.}
    \nonumber
    \\
    &+& \mu_S S_1' S_1^\dagger S_0^\dagger + \mu_1 \eta \phi  S_1^{'\dagger} + \mu_2 \eta^\dagger \phi S_0 + \text{h.c.}
    \\
    \nonumber
    &+& m_\eta^2 \eta^\dagger \eta + m_{S_0}^2 S_0^\dagger S_0 + m_{S_1}^2 S_1^\dagger S_1 + m_{S_1'}^2  S_1^{'\dagger} S_1' + ...\;,
\end{eqnarray}
where the term $\mu_S$ breaks softly the $\mathcal{Z}_2$ symmetry. The rest of the scalar potential is omitted for simplicity.
The effective Yukawa associated to the neutrino masses is given by
\begin{equation} \label{eq:yukawa nu 2}
    \left( Y_\nu \right)_{\alpha\beta} \approx \frac{1}{(16\pi^2)^2} \frac{\mu_S \mu_1 \mu_2}{M_i^3} \left[ \left( \Delta m_0^2 c^2_0 + \Delta m_+^2 s^2_+ \right) F^{(1)}_i + F^{(2)}_i \right]\, (Y_1)_{\alpha i} (Y_2)_{\beta i}.
\end{equation}
Here, $M_i$ are the mass eigenvalues of the vector-like fermions $E$, $\Delta m_0^2$ is the mass difference between the neutral eigenstates coming from the mixing of ($\eta^0$,$S_0$) with mixing angle $\cos \theta_0 \equiv c_0$ and $\Delta m_+^2$ the same for the charged eigenstates of ($\eta^+$,$S_1'$) with mixing angle $\sin \theta_+ \equiv s_+$. 

$F_{i}$ are the dimensionless loop integrals of the form,
\begin{small}\begin{subequations} \label{eq:F integrals 2}
    \begin{equation} %\label{eq:}
       F^{(1)}_{i} = M_i^6 \iint\limits_{(k,q)} \frac{ 1 }{ (k^2-M_i^2) (k^2-M_W^2) (k^2-m_{S_1}^2) (q^2-m_{\eta^+}^2) (q^2-m_{S_1}^2) ((k+q)^2-m_{\eta^0}^2) ((k+q)^2-m_{S_0}^2) },                                                                                                                                                                                     
    \end{equation}
    \begin{eqnarray} %\label{eq:}
        F^{(2)}_{i} & = & M_i^4 \iint\limits_{(k,q)} \frac{ 1 }{ (k^2-M_i^2) (k^2-M_W^2) (k^2-m_{S_1}^2) (q^2-m_{S_1}^2) ((k+q)^2-m_{S_0}^2) } \nonumber \\
        & & \qquad \qquad \left[ \frac{1}{(q^2-m_{\eta^+}^2)}  
        - \frac{1}{((k+q)^2-m_{\eta^0}^2)} \right].
    \end{eqnarray}
\end{subequations}\end{small}
Both integrals can be written in terms of simple one-loop and two-loop integrals, for which analytical expressions can be found, see Appendix~\ref{sec:appendix 2loop}.

In the same fashion as before, the neutrino mass scale is given in figure~\ref{fig:NuMass plot 2} in terms of the couplings of the model and the mass of $S_0$, keeping other masses of order $1$ TeV. Here, we consider no hierarchy in the Yukawas, obtaining a characteristic mass scale for neutrinos. The contributions for both integrals \eqref{eq:F integrals 2} are considered separately, plotted as dashed lines with labels $m_\nu^{(a)}$. The combination \eqref{eq:yukawa nu 2} is depicted as a solid line.

The overall behaviour is similar to figure~\ref{fig:NuMass plot 1}, but with a cancellation among diagrams. For small masses there is a visible suppression of the neutrino mass scale that even vanishes when $m_{S_0}^2 \approx m_{S_1'}^2$, leading to a lower neutrino mass compared to the previous example.

%% file: Section5_DM.tex
%%%%%%%%%%%%%%%%%%%%%%%%%%%%%%%%%%%%%%%%%%%%%%%%%%%%%%%%%%%%%%%%%%%%%%%%%%%%%%%%%%%%%%%%%%%%%%%
\section{Dirac Nature of Neutrinos and Dark Matter Stability}
\label{sec:dm}
%%%%%%%%%%%%%%%%%%%%%%%%%%%%%%%%%%%%%%%%%%%%%%%%%%%%%%%%%%%%%%%%%%%%%%%%%%%%%%%%%%%%%%%%%%%%%%

Before ending let us briefly discuss a very important and appealing feature of Dirac neutrino mass models, i.e. their possible connection with dark matter stability. It is clear that, as we discussed before in section ~\ref{sec:classification} and more specifically in section ~\ref{sec:models}, the Dirac nature of neutrinos implies the presence of a symmetry. It can be a new additional symmetry but it can also be the \sm $B-L$ or a residual subgroup of it \cite{Ma:2014qra,Ma:2015mjd,Ma:2015raa,Chulia:2016ngi,Bonilla:2018ynb,Ma:2019yfo}. Although it is natural to connect the lepton number conservation with the Dirac nature of neutrinos, the following arguments in this section are completely general and also cover symmetries not related with $B-L$ or lepton number. 

In order for the neutrinos to be Dirac in nature, neutrinos must transform non-trivially under the new symmetry. Assuming this scenario, after spontaneous symmetry breaking the Majorana mass terms should be forbidden while $\bar{\nu}_L \nu_R$ is allowed. It has been extensively studied in the literature that this same symmetry that protects the Dirac nature of neutrinos can, at the same time, be responsible of the stability of a dark matter candidate. Here, we want to explore this idea in more details starting from the simple models given in section ~\ref{sec:models}. Then we will generalize the core idea and show that it can be used in much broader scenarios. 

The two simple example models of section ~\ref{sec:models} make use of two discrete symmetries, of which $\mathcal{Z}_4$ plays the role of protecting the Diracness of neutrinos. Neutrinos transform as $z = e^{i \pi/2}=i$, automatically forbidding all Majorana mass terms for neutrino fields, since they are not $\mathcal{Z}_4$ invariant. This ensures that neutrinos are Dirac particles. 

Now to further see how in these models the dark matter stability can automatically come from the same $\mathcal{Z}_4$ symmetry, let us first introduce the concept of \textit{dark sector}. We define it as the set of particles which cannot decay into only Standard Model particles, irrespective of the masses of the particles involved. The lightest of the fields of the dark sector will then be automatically stable and thus a viable dark matter candidate \footnote{It is also possible to have multi-component dark matter in such a setup if there are more than one dark sector decoupled from the rest.}.

In both example models the dark sector consists of fermions which are even and scalars which are odd under the $\mathcal{Z}_4$ symmetry\footnote{We like to remind that odd(even) are fields which carry odd(even) powers of the $z = e^{i2\pi/4}$ charge, with $z^4 =1$.}. Note that all the \sm fermions are odd under $\mathcal{Z}_4$, i.e. they transform as $z^{2n+1}$, and all the vev carrying scalars are even, i.e. they transform as $z^{2n}$ with $n$ integer and $z^4 = 1$. Lorentz invariance forces the fermions to appear in pairs and thus any Lorentz invariant combination of \sm fermions will transform as even powers of $z$. This implies that all the odd scalars of the models will automatically be part of the dark sector, since $\mathcal{Z}_4$ is forbidding the decay of an odd scalar into a pair of fermions, as well as into even scalars. Similarly, any even fermion will also belong to the dark sector because it cannot decay to only the Standard Model. 

In conclusion, the lightest of the odd scalars and the even fermions will be stable as long as the $\mathcal{Z}_4$ symmetry is preserved. If one breaks the $\mathcal{Z}_4$ symmetry
of these models, then the Dirac nature of neutrinos and the stability of dark matter will be simultaneously lost. As a corollary, this implies that all vev-carrying scalars (including the \sm Higgs) always transform trivially under the conserved $\mathcal{Z}_n$ group and therefore will be even.

This same argument can be applied not only for $\mathcal{Z}_4$ but to any $\mathcal{Z}_{2n}$ or $U(1)$ symmetry groups. Since $\mathcal{Z}_2$ leads to Majorana neutrinos, the simplest group that simultaneously achieves Diracness plus stable dark matter with no accidental or explicit extra symmetries is $\mathcal{Z}_4$. 

\begin{figure}[h!]
\centering
    \includegraphics[width=0.5\textwidth]{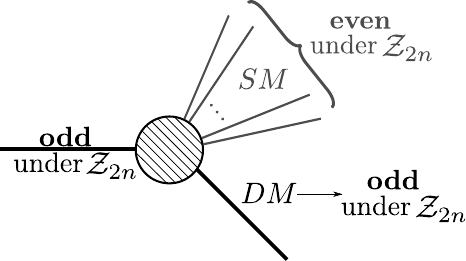} \, \, \,  
    \includegraphics[width=0.3\textwidth]{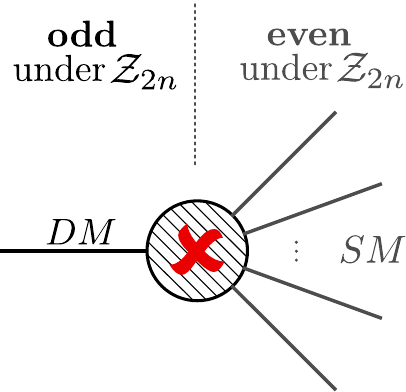}
    \caption{\footnotesize When the \sm fields transform as an even power of the group $\mathcal{Z}_{2n}$ then any odd particle will be automatically part of the dark sector.}
    \label{fig:oddsm}
\end{figure}

Using this idea as a guiding post we can flesh out an even simpler situation. If all the \sm particles transform as even powers of a $\mathcal{Z}_{2n}$, then automatically all the odd particles will be part of the dark sector, irrespective of the Lorentz nature of the fields involved. The powerful yet simple idea behind this mechanism is that an odd particle cannot decay into a combination of even particles. If neutrinos transform as even powers, then both $\mathcal{Z}_2$ and $\mathcal{Z}_4$ would lead to Majorana neutrinos, in accordance with \cite{Hirsch:2017col}. Therefore the simplest symmetry group that achieves Diracness and dark matter stability in this context would be $\mathcal{Z}_6$, with neutrinos transforming either as $\omega^2$ or $\omega^4$, with $\omega^6 = 1$, see \cite{Bonilla:2018ynb} for further details.  We can see this behavior graphically in figure~\ref{fig:oddsm}.

Note that we have only discussed general $U(1)$ or $\mathcal{Z}_{2n}$ groups. When the conserved group is an odd $\mathcal{Z}_{n}$ group, neutrinos will automatically be Dirac particles as discussed in \cite{Hirsch:2017col}. However in this case, the stability of dark matter is not straightforward. For a $\mathcal{Z}_{n}$ with $n$ a prime number one can find a dark matter decay channel due to the fact that the operator $\nu_R \nu_R$, singlet under the \sm, will transform non-trivially under the discrete symmetry. Taking advantage of the oddness of the group order, $(\nu_R \nu_R)^k$ can take any value depending on the value of $k$, see discussion in \cite{Bonilla:2018ynb, Bonilla:2019hfb} for more details.  

As a simple example of this idea let us take the conserved group to be $\mathcal{Z}_3$, $\nu_R \sim \omega = e^{i 2 \pi / 3}$. Then $\nu_R \nu_R \sim \omega^2$, $(\nu_R \nu_R)^2 \sim \omega$, $(\nu_R \nu_R)^3 \sim 1$. With this choice of charges any scalar $S$ will have an allowed effective decay operator of the form $S (\nu_R \nu_R)^k$ and any fermion $\Psi$ will have a decay channel of the form $\bar{\Psi} \nu_R (\nu_R \nu_R)^k$, with $k = 0, 1$ or $2$. Thus, in a UV complete theory where a renormalizable UV completion of these operators is possible, these particles cannot be part of the dark sector. This is because, they will decay into neutrinos and/or to the scalars out of the dark sector. 

Having said that, in such cases it is possible that the dark matter candidate is not absolutely stable, but it decays slowly enough so that its half life is much larger than age of the Universe. In such scenario, it can be a phenomenologically viable dark matter candidate. The other possibility is that the effective decay operators cannot be UV completed in the model due to its limited particle content. If this happens, then a new accidental symmetry will appear in such a model, protecting the dark matter against decaying. Explicit examples of such scenarios in the context of a $\mathcal{Z}_3$ symmetry have been discussed in \cite{Bonilla:2019hfb}. 

Finally, let us also mention the possible exception to the arguments against unsuitability of odd $\mathcal{Z}_n$ to, on their own, protect the dark matter stability. If the odd $\mathcal{Z}_n$ group of interest is a non-prime $\mathcal{Z}_n$ group, e.g. $\mathcal{Z}_9$, $\mathcal{Z}_{15}$, etc, with neutrinos transforming as multiples of the divisors of $n$, then we can have a completely stable dark matter candidate without an accidental symmetry. The simplest case of this type is $\mathcal{Z}_9$ group with the \sm particles transforming as $\omega^{3n}; \omega^9 = 1$. For such a scenario, any particle that does not transform as $\omega^{3n}$ will not decay to only \sm fields due to $\mathcal{Z}_9$. For example, an scalar transforming as $\omega$ will not be able to decay into any combination of \sm fermions, which will transform as $\omega^{3n}$, since $1+3n \neq 0$ modulo $9$ for any $n\in \mathbb{N}$. Further generalizations of this mechanism are also possible, but we will not go into details as the more complex generalizations will share the same key conceptual features discussed before. \\

We end this section by providing a summary of the above arguments:
\begin{itemize}
\item As we discussed in this section, the Dirac nature of neutrinos and dark matter stability can be intimately connected to each other with the same symmetry.
\item For an Abelian symmetry, it is possible to have Dirac neutrinos and stable dark matter without accidental symmetries if the remnant abelian group is of order non-prime, with the only exception of $\mathcal{Z}_2$. The $\mathcal{Z}_2$ group is too small to protect the Diracness of neutrinos. 
\item For the case of an Abelian discrete even $\mathcal{Z}_{2n}$ group, if all the \sm fields transform as even powers of the group, then any odd particle will be part of the dark sector and the lightest particle will be a good dark matter candidate.
\item For the case of an Abelian discrete even $\mathcal{Z}_{2n}$ group, when all the \sm fermions transform as odd powers while all the non-dark sector scalars are even, then all even fermions and odd scalars of the model will be part of the dark sector with the lightest of them being dark matter. Lorentz symmetry plays a fundamental role in this case.
\item For an Abelian discrete odd $\mathcal{Z}_{n}$ group with $n$ a prime number, it is not possible to have both Diracness of neutrinos and dark matter stability protected by only this symmetry. Nonetheless, in such cases the model can still have a phenomenologically viable decaying dark matter. Alternatively, as discussed before, the model might also have new accidental symmetries which can forbid the dark matter decay. 
\item These arguments can be easily generalized for other bigger groups and more involved scenarios.
\end{itemize}

As a conclusion, we can see that the symmetry which protects the Dirac nature of neutrinos can be related with dark matter stability in a plethora of different scenarios without the need to invoke any new explicit or accidental symmetries. Such models provide an attractive  possibility to have both Dirac neutrinos and dark matter using the same symmetry. Thus, this provides a interesting framework while being minimalistic in its symmetry content.

%% file: Section6_Summary.tex
%%%%%%%%%%%%%%%%%%%%%%%%%%%%%%%%%%%%%%%%%%%%%%%%%%%%%%%%%%%%%%%%%%%%%%%%%%%%%%%%%%%%%%%%%%%%%%%%%%%%%%%%
%%%%%%%%%%%%%%%%%%%%%%%%%%%%%%%%%%%%%%%%%%%%%%%%%%%%%%%%%%%%%%%%%%%%%%%%%%%%%%%%%%%%%%%%%%%%%%%%%%%%%%%%
\section{Discussion and Summary} \label{sec:summary}%%%%%%%%%%%%%%%%%%%%%%%%%%%%%%%%%%%%%%%%%%%%%%%%%%%%
%%%%%%%%%%%%%%%%%%%%%%%%%%%%%%%%%%%%%%%%%%%%%%%%%%%%%%%%%%%%%%%%%%%%%%%%%%%%%%%%%%%%%%%%%%%%%%%%%%%%%%%%
%%%%%%%%%%%%%%%%%%%%%%%%%%%%%%%%%%%%%%%%%%%%%%%%%%%%%%%%%%%%%%%%%%%%%%%%%%%%%%%%%%%%%%%%%%%%%%%%%%%%%%%%

We have discussed the complete decomposition of the Dirac neutrino mass operator $\bar{L} \phi^c \nu_R$ at two-loop order. We have identified all the 1PI topologies and diagrams with 3 external legs, two-loops and 3,4-point vertices which gives the dominant contribution to the neutrino mass. We call such diagrams as \textit{genuine}. From an initial set of 70 topologies, only 5 satisfy this genuineness criteria (figure~\ref{fig:topologies}), obtained after removing tadpoles, self-energy diagrams and non-renormalizable contributions.

A set of 18 renormalizable diagrams are generated straightforward from the 5 genuine topologies. We classify them in three different classes depending on the requirements imposed to their possible particle content in order to generate a genuine two-loop model. The three diagrams generated from topologies T1 and T2 given in figure~\ref{fig:diagrams 1} are genuine in general. This means that there is no particular field or symmetry breaking requirement in order for these diagrams to be the dominant contribution to neutrino masses. Meanwhile, the other 15 diagrams contain a one-loop realization of a fermion-fermion-scalar vertex (figure~\ref{fig:diagrams 2}) or a three scalar vertex (figure~\ref{fig:diagrams 3}). The former is genuine if one provides a symmetry transformation that forbids not only the tree-level but also the one-loop diagram and breaks it softly allowing the two-loop mass diagram. The latter always requires that the three scalars of the loop vertex are $\phi$, $\phi$ and $S\equiv(\mathbf{1},\mathbf{1},-1)$. The antisymmetric nature of $SU(2)_L$ contractions makes the local tree-level operator $\phi(x) \phi(x) S(x)$ zero but not its loop realizations, consequently forbidding the reduction of this class of two-loop diagrams into their corresponding one-loop diagrams (see figure~\ref{fig: example HHS} for details).

We have found that every case of neutrino mass generation from the operator $\bar{L} \phi^c \nu_R$ at two-loops can be written in terms of 6 mass diagrams or integrals (figure~\ref{fig:massdiagrams}). These integrals can be decomposed in terms of two master integrals for which analytical expressions already exist \cite{Sierra:2014rxa,Martin:2016bgz}.

We have shown how one can generate models from our classification, listing all possible \sm quantum numbers with $SU(2)_L$ representations up to triplets. Although, for simplicity,  we only discussed the cases with colour singlet fields, nevertheless as explained before, introducing non-trivial representations of $SU(3)_C$ is straightforward as the external fields are colour blind. To illustrate how our classification can be used to generate genuine models, we have constructed and discussed in detail two different Dirac neutrino mass models. Each of the models are built from two characteristic sets of diagrams explained in the previous section. One of the example models uses a completely genuine topology, so that the two-loop contribution is guaranteed to be the leading order contribution to neutrino masses. The second example model makes use of the non-locality of the operator $\phi \phi S$ in order to be non-reducible. We have shown that these type of models are able to fit neutrino oscillation data for reasonable values of the masses and parameters. Such models are testable and a part of the parameter space is already excluded by collider searches. In this direction, a more involved analysis of the phenomenology of these models would be needed for detailed quantitative results. 

Finally, we have treated the connection between the symmetry that protects the Dirac nature of neutrinos and the stability of dark matter. We have shown that for Dirac neutrino mass models, the symmetry protecting the Diracness of neutrinos can also simultaneously protect the dark matter from decay. We showed that this relation holds true in general for a wide class of symmetries. We also discussed the various possibilities and classes given a particular set of symmetries and the transformation of the \sm particles under it. The relationship between the Diracness of neutrinos and dark matter stability is an attractive possibility intimately connecting the neutrino and dark matter physics through the same symmetry.

%% file: Appendix_QuantumNumbers.tex
%%%%%%%%%%%%%%%%%%%%%%%%%%%%%%%%%%%%%%%%%%%%%%%%%%%%%%%%%%%%%%%%%%%%%%%%%%%%%%%%%%%%%%%%%%%%%%%%%%%%%%%%
%%%%%%%%%%%%%%%%%%%%%%%%%%%%%%%%%%%%%%%%%%%%%%%%%%%%%%%%%%%%%%%%%%%%%%%%%%%%%%%%%%%%%%%%%%%%%%%%%%%%%%%%
\section{Quantum numbers of the internal fields} \label{sec:appendix qn}
%%%%%%%%%%%%%%%%%%%%%%%%%%%%%%%%%%%%%%%%%%%%%%%%%%%%%%%%%%%%%%%%%%%%%%%%%%%%%%%%%%%%%%%%%%%%%%%%%%%%%%%%
%%%%%%%%%%%%%%%%%%%%%%%%%%%%%%%%%%%%%%%%%%%%%%%%%%%%%%%%%%%%%%%%%%%%%%%%%%%%%%%%%%%%%%%%%%%%%%%%%%%%%%%%

In this appendix, we give the SM quantum numbers for the diagrams $D_2^X - D_5^X$ in figure~\ref{fig:diagrams-models-naming}. Every table in this section obeys the same convention as table~\ref{tab:QN 1} for $D_1^X$: (1) two tables are given for $SU(2)_L$ representations and hypercharge, (2) we consider representations up to triplets, and (3) two input fields are needed $X_1$ and $X_2$ with hypercharges $\alpha_1$ and $\alpha_2$, respectively, while the $SU(2)_L$ representations are explicitly given in the first row and column of each table. Further explanations in section~\ref{sec:generating models}.

\begin{table}[h!]
\centering
    \begin{tabular}{|*{4}{c|}}
        \hline\hline
        \multicolumn{4}{c}{\textbf{Hypercharge of $X_6$ for $D_{2-4}^X$}} \\
        \hline
         & \makebox[7em]{$D_2^X$} & \makebox[5em]{$D_3^X$} & \makebox[3em]{$D_4^X$} \\
        \hline
       \makebox[3em]{$X_6$} & $-\alpha_1-\alpha_2-1/2$ & $\alpha_2+1/2$ & $\alpha_1$ \\
        \hline
    \end{tabular}
    \\[3ex]
    \begin{tabular}{|c||*{3}{c|}|*{3}{c|}|*{3}{c|}}
        \hline\hline
        \multicolumn{10}{c}{\textbf{$SU(2)_L$ representations of $X_6$ for $D_{2-4}^X$}} \\
        \hline
        \backslashbox[3em]{$X_2$}{$X_1$} & \multicolumn{3}{c||}{1} & \multicolumn{3}{c||}{2} & \multicolumn{3}{c|}{3} \\
        \hline\hline
        & \multicolumn{3}{c||}{$X_6$} & \multicolumn{3}{c||}{$X_6$} & \multicolumn{3}{c|}{$X_6$} \\
        \hline
        & $D_2^X$ & $D_3^X$ & $D_4^X$ & $D_2^X$ & $D_3^X$ & $D_4^X$ & $D_2^X$ & $D_3^X$ & $D_4^X$ \\
        \hline\hline
        1 & 2 & 2 & \slashbox[2.4em]{1}{3} & \slashbox[2.4em]{1}{3} & 2 & 2 & 2 & 2 & \slashbox[2.4em]{1}{3} \\
        \hline
        2 & \slashbox[2.4em]{1}{3} & \slashbox[2.4em]{1}{3} & \slashbox[2.4em]{1}{3} & 2 & \slashbox[2.4em]{1}{3} & 2 & \slashbox[2.4em]{1}{3} & \slashbox[2.4em]{1}{3} & \slashbox[2.4em]{1}{3} \\
        \hline
        3 & 2 & 2 & \slashbox[2.4em]{1}{3} & \slashbox[2.4em]{1}{3} & 2 & 2 & 2 & 2 & \slashbox[2.4em]{1}{3} \\
        \hline
    \end{tabular}
\caption{Standard Model quantum numbers for the field $X_6$ of the diagrams $D_2^X$, $D_3^X$ and $D_4^X$ in figure~\ref{fig:diagrams-models-naming}. Two input fields are needed $X_1$ and $X_2$ with hypercharges $\alpha_1$ and $\alpha_2$, respectively, and $SU(2)_L$ representations explicitly given in the first row and column of the right table. These tables should be completed with table~\ref{tab:QN 1} which contains the quantum numbers for the fields $X_3$, $X_4$ and $X_5$, common for all the diagrams. For simplicity, all the fields are colour singlets.}
\label{tab:QN 2-4}
\end{table}

For simplicity we do not give one set of tables for every diagram, but we unified the tables of $D_{2-4}^X$ with that of $D_1^X$ (table~\ref{tab:QN 1}). From figure~\ref{fig:diagrams-models-naming} it can be shown that the diagram $D_1^X$ can be obtained by shrinking the field $X_6$. This means that for the diagrams $D_2^X$, $D_3^X$ and $D_4^X$ the fields are identical to those of $D_1^X$, except for $X_6$. For each assignment of $SU(2)_L$ representation and hypercharge of the fields $X_1$ and $X_2$, the quantum numbers of $X_6$ for each diagram in $D_{2-4}^X$ are depicted table~\ref{tab:QN 2-4}, completing the charge assignment for fields $X_{1-5}$ in table~\ref{tab:QN 1}, identical for all the diagrams.

The only diagram that do not shrink to $D_1^X$ is $D_5^X$, for which a specific set of tables is needed. For $D_5^X$ the quantum numbers are given in table~\ref{tab:QN 5}, in the same fashion as the example already discussed.

\begin{table}
\centering
    \begin{tabular}{|*{6}{c|}}

        \hline\hline

        \multicolumn{6}{c}{\textbf{Hypercharge for $D_5^X$}} \\

        \hline

        \makebox[3em]{$X_1$} & \makebox[3em]{$X_2$} & $X_3$ & \makebox[3em]{$X_4$} & \makebox[3em]{$X_5$} & \makebox[3em]{$X_6$} \\

        \hline

        $\alpha_1$ & $\alpha_2$ & $-\alpha_1+1/2$ & $\alpha_1$ & $-\alpha_1-\alpha_2+1/2$ & $\alpha_1+\alpha_2$ \\

        \hline
    \end{tabular}
    \\[3ex]
    \begin{tabular}{|c||*{4}{c|}|*{4}{c|}|*{4}{c|}}
        \hline\hline
        \multicolumn{13}{c}{\textbf{$SU(2)_L$ representations for $D_5^X$}} \\
        \hline
        \backslashbox[3em]{$X_2$}{$X_1$} & \multicolumn{4}{c||}{1} & \multicolumn{4}{c||}{2} & \multicolumn{4}{c|}{3} \\
        \hline\hline
        & \makebox[2em]{$X_3$} & \makebox[2em]{$X_4$} & \makebox[2em]{$X_5$} & \makebox[2em]{$X_6$} & \makebox[2em]{$X_3$} & \makebox[2em]{$X_4$} & \makebox[2em]{$X_5$} & \makebox[2em]{$X_6$} & \makebox[2em]{$X_3$} & \makebox[2em]{$X_4$} & \makebox[2em]{$X_5$} & \makebox[2em]{$X_6$} \\
        \hline\hline
        1 & 2 & 1 & 2 & 1 & \slashbox[2.4em]{1}{3} & 2 & \slashbox[2.4em]{1}{3} & 2 & 2 & 3 & 2 & 3 \\
        \hline
        2 & 2 & 1 & \slashbox[2.4em]{1}{3} & 2 & \slashbox[2.4em]{1}{3} & 2 & 2 & \slashbox[2.4em]{1}{3} & 2 & 3 & \slashbox[2.4em]{1}{3} & 2 \\
        \hline
        3 & 2 & 1 & 2 & 3 & \slashbox[2.4em]{1}{3} & 2 & \slashbox[2.4em]{1}{3} & 2 & 2 & 3 & 2 & \slashbox[2.4em]{1}{3} \\
        \hline
    \end{tabular}
\caption{Standard Model quantum numbers for the diagram $D_5^X$ (T5-vii and T5-x) in figure~\ref{fig:diagrams-models-naming}. We follow Tabs.~\ref{tab:QN 1} and \ref{tab:QN 2-4} with two input fields $X_1$ and $X_2$ with general hypercharges and $SU(2)_L$ representations up to triplets. All the fields are colour singlets.}
\label{tab:QN 5}
\end{table}

%% file: Appendix_2loopIntegrals.tex
\section{Computation of two-loop integrals} \label{sec:appendix 2loop}
%%%%%%%%%%%%%%%%%%%%%%%%%%%%%%%%%%%%%%%%%%%%%%%%%%%%%%%%%%%%%%%%%%%%%%%%%%%%%%%%%%%%%%%%%%%%%%%%%%%%%%%%
%%%%%%%%%%%%%%%%%%%%%%%%%%%%%%%%%%%%%%%%%%%%%%%%%%%%%%%%%%%%%%%%%%%%%%%%%%%%%%%%%%%%%%%%%%%%%%%%%%%%%%%%

In this section, we summarize the main tools needed in order to write every two-loop integral in terms of two master integrals. Two-loop integrals have been evaluated before in the literature and here we will follow \cite{vanderBij:1983bw,Sierra:2014rxa,Martin:2016bgz}. 

To illustrate how the decomposition of two-loop integrals into an analytic expression works, we take the loop functions \eqref{eq:F integrals 1} from the first example in section~\ref{sec:models}. Rewriting them into explicitly dimensionless integrals we get,
\begin{subequations} \label{eq:appendix F}
    \begin{equation} %\label{eq:}
        F^{(1)} = \iint\limits_{(k,q)} \frac{ 1 }{ (k^2-x_1) (k^2-x_2) (q^2-x_3) (q^2-x_4) (q^2-x_5) ((k+q)^2-1) }
    \end{equation}
    \begin{equation} %\label{eq:}
        F^{(2)} = \iint\limits_{(k,q)} \frac{ k \cdot q }{ (k^2-x_1) (k^2-x_2) (q^2-x_3) (q^2-x_4) (q^2-x_5) ((k+q)^2-1) },
    \end{equation}
\end{subequations}
with the following definitions,
\begin{equation} %\label{eq:}
    x_1 = \frac{m_\eta^2}{m_{S_3}^2}, \quad x_2 = \frac{M_{Ei}^2}{m_{S_3}^2}, \quad x_3 = \frac{M_{Nj}^2}{m_{S_3}^2}, \quad x_4 = \frac{m_{S_1}^2}{m_{S_3}^2}, \quad x_5 = \frac{m_{S_2}^2}{m_{S_3}^2},
\end{equation}
and
\begin{equation} %\label{eq:}
    \int\limits_k \equiv (16\pi^2)\int \frac{d^4\!k}{(2\pi)^4}.
\end{equation}

By the use of partial fraction, when various propagators have the same momenta, the integral can be write as a sum over integrals with less denominators,
\begin{equation} %\label{eq:}
    \frac{1}{(k^2-x_1)(k^2-x_2)} = \frac{1}{x_1-x_2} \left( \frac{1}{k^2-x_1} - \frac{1}{k^2-x_2} \right).
\end{equation}
Moreover, integrals with momenta in the numerator which coincides with that of one of the propagators can be reduced as,
\begin{equation} %\label{eq:}
    \frac{q^2}{(k^2-x_1)(q^2-x_2)} = \frac{1}{k^2-x_1} + \frac{x_2}{(k^2-x_1)(q^2-x_2)}.
\end{equation} 
Making use of only these two expressions one can write every two-loop integral in terms of the basis,
\begin{subequations}
    \begin{equation} %\label{eq:}
        \mathbf{A}(x) = \int\limits_k \frac{1}{k^2-x},
    \end{equation}
\vspace*{-0.5cm}
    \begin{equation} %\label{eq:
        \mathbf{I}(x,y,z) = \iint\limits_{(k,q)} \frac{1}{(k^2-x)(q^2-y)((k+q)^2-z)},
    \end{equation}
\end{subequations}
for which analytical expression can be easily found in the literature, see for example \cite{Passarino:1978jh,Martin:2016bgz}.

Particularly, for the two-loop integrals given in \eqref{eq:appendix F} and used for the numerical analysis of figure~\ref{fig:NuMass plot 1}, the decomposition in terms of the master integrals $\mathbf{A}$ and $\mathbf{I}$ is  
\begin{small}
\begin{equation} %\label{eq:}
  \hspace*{-0.5cm}
  F^{(1)} = \frac{1}{(x_1-x_2)(x_3-x_4)} \left\lbrace \frac{1}{x_3-x_5} \left[ \mathbf{I}(x_1,x_3,1) - \mathbf{I}(x_1,x_5,1) - \mathbf{I}(x_2,x_3,1) + \mathbf{I}(x_2,x_5,1) \right] - \left( x_3 \leftrightarrow x_4 \right) \right\rbrace,
\end{equation}
\end{small}
and
\begin{small}
\begin{dmath} %\label{eq:}
  % \hspace*{-1cm}
        F^{(2)} = \frac 12 (1-x_2-x_5) F^{(1)} 
        \\
        + \frac{1}{(x_1-x_2)(x_3-x_4)} \left\lbrace \frac{1}{x_3-x_5} \left[ \mathbf{A}(x_1)\mathbf{A}(x_3) - \mathbf{A}(x_1)\mathbf{A}(x_5) - \mathbf{A}(x_2)\mathbf{A}(x_3) + \mathbf{A}(x_2)\mathbf{A}(x_5)
         \\
         - (x_1-x_2)( \mathbf{I}(x_1,x_3,1) - \mathbf{I}(x_1,x_5,1) ) \right] - \mathbf{I}(x_1,x_3,1) + \mathbf{I}(x_2,x_3,1)
         \\
         - \left( x_3\nolinebreak \leftrightarrow x_4 \right) \right\rbrace,
\end{dmath}
\end{small}
where we have used that $k \cdot q = \frac 12 \left[ (k+q)^2-k^2-q^2 \right]$.

The decompositions for all the diagrams in figure~\ref{fig:massdiagrams} in terms of the master integrals can be found in \cite{Sierra:2014rxa}.